\documentclass[runningheads]{llncs}
\usepackage[T1]{fontenc}
\usepackage{graphicx}
\usepackage{booktabs}
\usepackage[misc]{ifsym}

\newcommand{\corr}{(\Letter)}
\usepackage{mwe}
\usepackage{multirow}
\usepackage{subcaption} 
\usepackage{amssymb}
\usepackage{enumitem}
\usepackage{xcolor}
\usepackage[final]{microtype}
\usepackage{hyperref}
\usepackage{amsmath}
\DeclareRobustCommand{\method}{\texttt{AE-PSL}}
\DeclareRobustCommand{\cls}{\texttt{[CLS]}}
\definecolor{DeltaBlue}{HTML}{3B6EA8}
\definecolor{dblue}{HTML}{3B6EA8}
\newcommand{\abres}[3]{$#1 \pm #2$\,{\scriptsize\color{DeltaBlue}$(#3)$}}
\newcommand{\abresbest}[3]{$\mathbf{#1 \pm #2}$\,{\scriptsize\color{DeltaBlue}$\mathbf{(#3)}$}}

\begin{document}

\title{AutoEncoder-Compressed Parallel Split Learning for Pre-trained Model Fine-Tuning}

\titlerunning{AutoEncoder-Compressed Parallel Split Learning}

\author{Author information scrubbed for double-blind reviewing}
\author{Bas Meuwissen \corr \and Vasileios Tsouvalas \and Nirvana Meratnia}
\authorrunning{B. Meuwissen et al.}
\institute{Eindhoven University of Technology, Eindhoven, The Netherlands \\
\email{b.meuwissen@student.tue.nl, \{v.tsouvalas,n.meratnia\}@tue.nl}
}
\maketitle              

\begin{abstract}
Distributed Fine-Tuning (DFT) of large-scale Foundation Models (FMs) on resource-constrained edge devices is limited by local compute constraints and communication overhead. Parallel Split Learning (PSL) reduces client-side computation by keeping few model layers on each client and offloading the remaining computation to the server; however, clients must exchange intermediate activations and gradients with the server at every training step. Existing SL communication-compression methods mainly rely on task-agnostic heuristics, such as sparsification and quantization. While learnable SL compressors can better adapt to intermediate representations, they require co-training with the target model. Therefore, directly inserting them into off-the-shelf FMs introduces feature-distribution misalignment and degrades DFT performance. To address this, we propose \method{}\footnote{Public code repository: \url{https://github.com/Nousphera/AE_PSL}}, a communication-efficient PSL framework that compresses intermediate activations and gradients using a lightweight AutoEncoder (AE) placed at the split layer. To ensure compatibility of AE compression with pre-trained FMs, \method\ introduces a novel two-stage alignment mechanism, which adapts the AE to the pre-trained model's feature manifold and client-specific feature distributions before DFT. 
We evaluate \method\ across four vision datasets and various communication reduction settings, showing that it preserves downstream accuracy at $10.2\times$ reduction; at the same communication budget, the strongest evaluated heuristic baseline trails \method~by $5.2$ percent point.
This demonstrates that \method\ offers a stronger accuracy--communication trade-off than heuristic SL compression.


\vspace{-3pt}
\end{abstract}

\vspace{-15pt}  \section{Introduction} \vspace{-3pt} 

Large-scale transformer-based Foundation Models (FMs) have established state-of-the-art performance across different modalities and complex tasks. These models are trained on a massive amount of data gathered on a central server. Meanwhile, the proliferation of Internet of Things (IoT) devices has led to a vast amount of data residing at the edge. Such data can be valuable for fine-tuning FMs, yet centralizing it is often infeasible due to privacy constraints, while local fine-tuning is limited by the memory and compute demands of large models. \par

To address these constraints, distributed learning schemes have been explored. Federated Learning (FL)~\cite{mcmahan2017FL} enables collaborative FM fine-tuning while keeping client data localized; yet it requires each client to train the full model, imposing substantial storage and computational burdens on edge devices. Split Learning (SL)~\cite{gupta2018SL} alleviates this by splitting the model at a cut layer into client-side and server-side sub-models, allowing clients to execute only the \textit{lightweight} client-side component. Recent Parallel SL (PSL) variants, such as Scalable Aggregated SL (SASL)~\cite{lyu2023SASL}, further improve SL’s scalability through parallel client training and reduced server-side computation. However, unlike FL’s round-based model-update exchange, SL requires clients to transmit intermediate activations and receive their corresponding gradients at every training iteration. This shifts the main cost from computation to communication, especially in transformer-based models whose intermediate tensors scale with sequence length~\cite{Zhao2023LLM-survey,singh2019FLvsSL}. \par

Existing SL communication-compression methods mostly rely on heuristics, such as quantization \cite{cohen2021sl-quant,oh2025adaptive_quant} and pruning \cite{castiglia2022vfl-topk,zheng2023rand-topk,alvetreti2025attention-dc}, which are simple but task-agnostic and can discard information critical to pre-trained FM representations. Learnable compression methods, including AutoEncoder (AE)-based modules, have shown promise in intermediate representation compression~\cite{eshratifar2019bottlenet,matsubara2022bottlefit,shao2020bottlenet,ayad2021slAE,guo2024split_medical,shiranthika2024splitfedzip,chen2025heterosfl,mudvari2024deprune}. However, prior AE-based methods target split-computing inference or jointly train the model with the AE module, making them ill-suited for fine-tuning off-the-shelf pre-trained FMs, where a randomly-initialized AE can distort intermediate representations and degrade performance. 
\par

To bridge this gap, we introduce \method, a scalable PSL framework for communication-efficient fine-tuning of off-the-shelf pre-trained FMs. \method~places a lightweight AE module at the split layer, with its encoder on the client and the decoder on the server, reducing the bidirectional communication overhead of activations and gradients. To stabilize fine-tuning and to avoid  representation distortion of FMs, we introduce an initial two-stage alignment process that adapts the AE module to general and client-specific feature distributions, without re-training the original model from scratch or centralizing private data. 
Our contributions are as follows:

\vspace{-3pt}
\begin{itemize}
    \item We introduce \method, an AE-compressed PSL framework for fine-tuning pre-trained vision FMs, reducing both activation and gradient communication through client-side encoders and a shared server-side decoder.
    \item We design a $2$-stage alignment protocol, comprising General Alignment (GA) and Client-Specific Alignment (CSA), that adapts the AE module to the pre-trained model and client-specific feature distributions, enabling stable integration with off-the-shelf FMs without centralizing private data.
    \item We evaluate \method\ across four vision datasets, showing that feature-aligned learnable compression provides a more robust accuracy--communication trade-off than heuristic SL methods, yielding $\sim10\times$ communication reduction with no accuracy degradation and negligible client-side computational overhead.

\end{itemize}

\vspace{-5pt} \section{Related Work} \label{sec:rw} \vspace{-3pt} 

\noindent \textbf{Distributed Training and PEFT.} Efficient FM adaptation on edge devices requires reducing client-side computation while avoiding centralized access to private data. FL preserves data locality, but requires clients to train the full model, imposing prohibitive memory and compute demands on resource-constrained devices~\cite{mcmahan2017FL}. SL addresses this by splitting the model between clients and a server~\cite{gupta2018SL}, while scalable variants such as SASL further enable parallel client training and reduce server-side backpropagation to $\mathcal{O}(1)$ through global loss aggregation~\cite{lyu2023SASL}. Orthogonally, PEFT methods such as LoRA~\cite{hu2022LoRA} have been integrated into Split FL (SFL) to reduce trainable parameters and weight-updates overhead during fine-tuning~\cite{lin2024splitlora,zhang2024hierarch_splitllm,chen2025privacy_sfl_lora,li2025ee_sl_lora,zhao2025SflLLM,lin2025hsplitlora,ma2025splitfrozen,Zhang2025SplitFT}. However, PEFT primarily targets model-parameter efficiency and does not reduce the per-iteration exchange of intermediate activations and gradients, which remains a major communication cost in transformer-based SL~\cite{Zhao2023LLM-survey,singh2019FLvsSL,fudala2025MPSL}. \method\ addresses this complementary cost by compressing intermediate representations within scalable PSL, while retaining split execution and PEFT-based fine-tuning.

\vspace{3pt} \noindent \textbf{Communication Compression in SL.} In SL, communication reduction for intermediate representations typically relies on heuristic or learned mechanisms. \textit{Heuristic compression} approaches rely on predefined, non-parametric operations such as quantization \cite{cohen2021sl-quant} or sparsification (e.g. top-$k$) \cite{castiglia2022vfl-topk}. To enhance convergence, Rand-Top-$K$ \cite{zheng2023rand-topk} extends standard top-$k$ selection by randomly transmitting a fraction of tokens outside the top-$k$ threshold. Attention-based Double Compression (ADC) \cite{alvetreti2025attention-dc} performs batch-wise compression by clustering sample activations with $K$-means using class-token attention scores, averaging activations within each cluster, and then retaining only the top-$k$ spatial tokens most aligned with each cluster centroid. C3-SL \cite{hsieh2022c3-sl} relies on algebraic compression, using circular convolution and superposition to encode batches of high-dimensional intermediate features into compact vectors that the cloud server reconstructs via circular correlation with fixed quasi-orthogonal random keys.

Although heuristic compression methods are simple and effective at moderate reduction levels, their fixed, task-agnostic rules cannot adapt to the structure of intermediate representations and may discard features that are important for downstream prediction. This motivates \textit{learnable compression}, which trains the encoding scheme to preserve task-relevant information under communication constraints. BottleNet \cite{eshratifar2019bottlenet} uses a convolutional AE-based architecture at the cut layer to reduce intermediate image features, while BottleNet++ \cite{shao2020bottlenet} extends this design with joint source–channel coding to improve robustness under unreliable communication. BottleFit \cite{matsubara2022bottlefit} replaces selected layers of a pre-trained model with a bottleneck module and uses a two-stage training procedure to first learn compressed representations and then adapt downstream layers to them. These methods focus on \textit{split computing}, where only forward inference is partitioned, rather than SL, where compressed intermediate representations must support end-to-end training across the cut layer. Pioneering works implement learnable compression in SL, however these require joint training from scratch and are designed for convolutional networks \cite{chen2025heterosfl,guo2024split_medical,ayad2021slAE,mudvari2024deprune,shiranthika2024splitfedzip}.
We address these limitations with an AE-based mechanism tailored to pre-trained feature distributions, enabling communication-efficient fine-tuning of FMs within scalable, parallel, label-private SL regimes.

\vspace{-5pt} \section{Background} \vspace{-3pt} \label{sec:background}

While SFL parallelizes client-side training, it imposes an $\mathcal{O}(n)$ computational and memory burden on the server, which must maintain distinct per-client sub-models and execute separate backpropagation passes. Scalable Aggregated Split Learning (SASL)~\cite{lyu2023SASL} resolves this bottleneck by reducing server-side backpropagation complexity to $\mathcal{O}(1)$ via global loss aggregation. Under the SASL framework, clients forward their local intermediate activations to a single, shared server model. The server processes these activations during the forward pass and returns predictions to the clients. Crucially, clients compute their local losses using private labels and transmit only the scalar loss values back to the server. The server aggregates these individual losses into a single global loss function to execute a single backpropagation pass. The resulting cut-layer gradients are then distributed back to the clients to complete local backpropagation. This workflow eliminates per-client server-side sub-models while maintaining label privacy.

\begin{figure*}[t]
    \includegraphics[width=\linewidth]{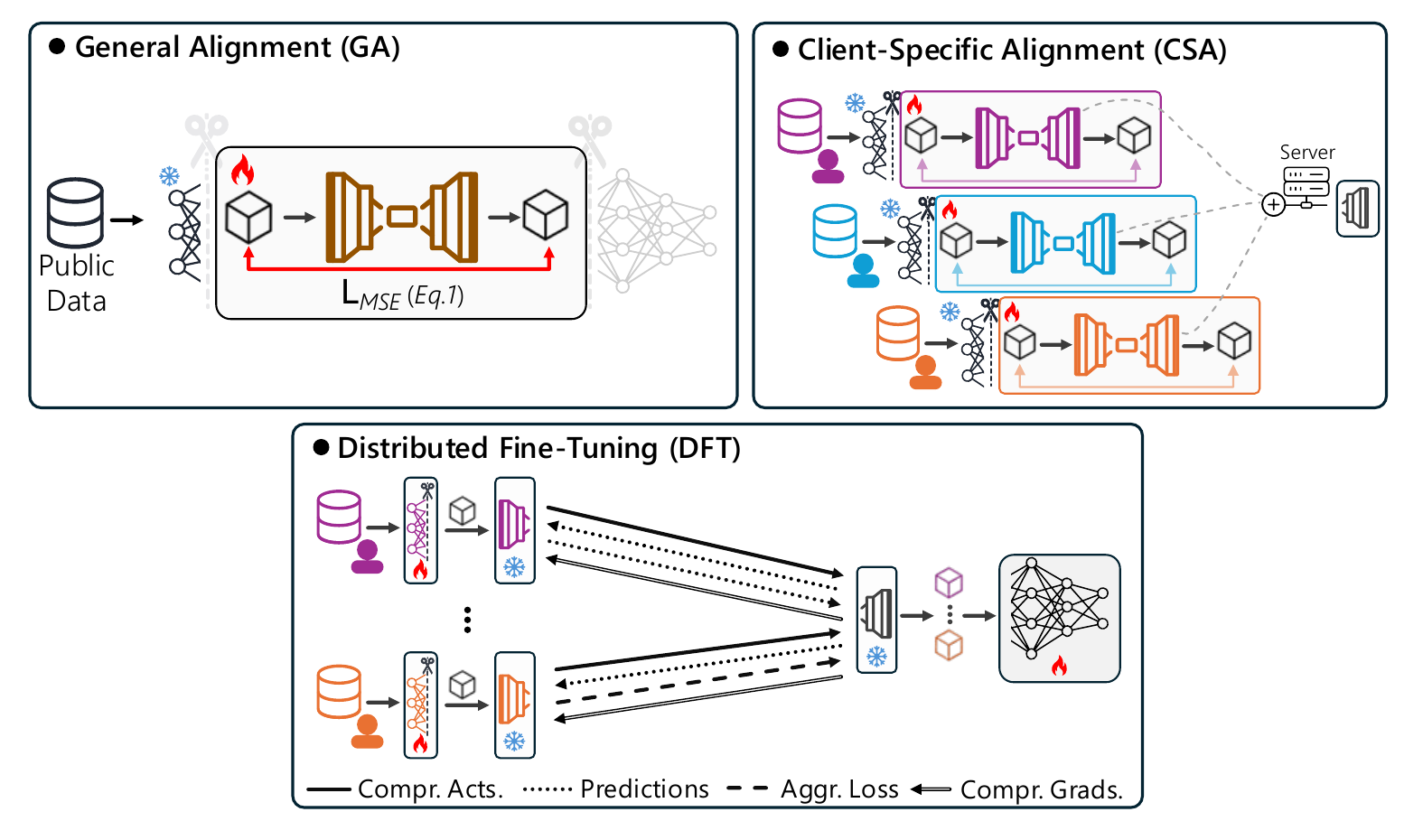}
    \caption{Overview of \method{}, consisting of three sequential stages: (1) \textit{GA} trains the AE on public data to match the pre-trained model's feature manifold. (2) \textit{CSA} locally adapts the AE to client-specific data, aggregating only the decoders at the server. (3) \textit{DFT} executes distributed training with frozen AEs, minimizing communication overhead by transmitting compressed activations and gradients.}
    \label{fig:overview}
\end{figure*}

\vspace{-5pt} \section{\method: Communication-efficient PSL via AutoEncoder} \vspace{-3pt}

To mitigate SL communication bottlenecks, we propose an AE-based module that compresses intermediate activations and gradients in PSL.

\vspace{5pt} \noindent \textbf{Notation.} We denote scalars by $x$, vectors by $\mathbf{x}$, matrices by $\mathbf{X}$, sets by $\mathcal{X}$, and neural networks by $f_{\boldsymbol{\theta}}$, where $\boldsymbol{\theta}$ denotes their parameters. Let $\mathcal{N}$ be a set of $N$ clients, each with local dataset $\mathcal{D}_n$ from which mini-batches $\mathcal{B}_n \subset \mathcal{D}_n$ are sampled, and let $\mathcal{B} = \bigcup_{n \in \mathcal{N}} \mathcal{B}_n$ denote the global batch.
A pre-trained transformer $f_{\boldsymbol{\theta}}$ is split at layer $s$ into client-side and server-side components $f_{\boldsymbol{\theta}_c}$ and $f_{\boldsymbol{\theta}_s}$. For client $n$, the client-side model outputs smashed data $(\mathbf{h}_{\mathrm{cls},n}, \mathbf{H}_n)$, consisting of the \cls\ activation and patch-token matrix, respectively. \par

\vspace{5pt} \noindent \textbf{Problem Formulation.} To reduce communication, AE-PSL inserts a learnable AE at the split point: the client-side encoder $E_{\boldsymbol{\phi}}$ compresses $\mathbf{H}_n$ from hidden dimension $d$ to latent dimension $d_z \ll d$, producing $\mathbf{Z}_n = E_{\boldsymbol{\phi}}(\mathbf{H}_n)$, while the server-side decoder $D_{\boldsymbol{\psi}}$ reconstructs $\widehat{\mathbf{H}}_n = D_{\boldsymbol{\psi}}(\mathbf{Z}_n)$. Our objective is to fine-tune $(f_{\boldsymbol{\theta}_c}, f_{\boldsymbol{\theta}_s})$ while preserving task-relevant information. To minimize communication overhead during this process, we transmit the compressed representations $\mathbf{Z}_n$ and their gradients instead of the raw intermediate activations $\mathbf{H}_n$.

\vspace{-5pt} \subsection{Two-Stage AutoEncoder Alignment} \label{sec:two-stage-AEA} \vspace{-3pt} 

A key challenge in using AEs with pre-trained models is avoiding catastrophic degradation caused by feature-distribution misalignment~\cite{eshratifar2019bottlenet}. We address this through an initial two-stage alignment procedure, aligning the AE to client feature distributions while requiring no server-side access to private downstream data.

\vspace{-5pt} \subsubsection{General Alignment (GA).} \label{sec:GA} \vspace{-3pt} 

We first insert the AE at the cut layer $s$ of the pre-trained model $f_{\boldsymbol{\theta}}$ and train it on the server-side using publicly available data (e.g., ImageNet~\cite{imagenet}), $\mathcal{D}_{\mathrm{p}}$, to reconstruct split-layer activations, while keeping $f_{\boldsymbol{\theta}_c}$ frozen. Specifically, we minimize the reconstruction MSE:

\vspace{-3pt}
\begin{equation} \label{eq:ga}
\resizebox{0.75\linewidth}{!}{$
(\boldsymbol{\phi}^{\mathrm{GA}}, \boldsymbol{\psi}^{\mathrm{GA}}) = \underset{\boldsymbol{\phi},\boldsymbol{\psi}}{\arg\min}\; \frac{1}{|\mathcal{D}_{\mathrm{p}}|} \sum_{\mathbf{x} \in \mathcal{D}_{\mathrm{p}}} \left\|\mathbf{H}(\mathbf{x}) - D_{\boldsymbol{\psi}}\!\left(E_{\boldsymbol{\phi}}(\mathbf{H}(\mathbf{x}))\right) \right\|_{2}^{2}~,%
$}
\end{equation}
\vspace{-4pt}


\noindent where $\mathbf{H}(\mathbf{x})$ denotes the patch-token activations extracted at split layer $s$ from sample $\mathbf{x}$, with $f_{\boldsymbol{\theta}_c}$ frozen. Thus, GA is a one-time alignment step for a given pre-trained model that adapts the AE to the model's feature manifold.

\vspace{-5pt} \subsubsection{Client-Specific Alignment (CSA).} \label{sec:CSA} \vspace{-3pt} 

While GA aligns the AE with the feature manifold of the pre-trained model, local client data can induce task-specific and non-independent and identically (non-IID) distributed feature shifts~\cite{FL_nonIID}. To this end, each client $n$ receives the GA-initialized pair $(\boldsymbol{\phi}^{\mathrm{GA}}, \boldsymbol{\psi}^{\mathrm{GA}})$ and locally warms up the AE on $\mathcal{D}_n$ using the same MSE reconstruction objective as in Eq.~\ref{eq:ga}, while keeping $f_{\boldsymbol{\theta}_c}$ frozen. Crucially, CSA is fully local and lightweight, requiring only the frozen client-side partition and AE module, with no forward passes over the $f_{\boldsymbol{\theta}_s}$ sub-model.

To preserve scalability, we avoid per-client server decoders. Instead, after CSA, each client keeps its adapted encoder $\boldsymbol{\phi}_n$ locally and sends only $\boldsymbol{\psi}_n$ to the server, which aggregates them by weight averaging (i.e., $\boldsymbol{\psi}^{\mathrm{global}} = \frac{1}{N}\sum_{n \in \mathcal{N}} \boldsymbol{\psi}_n$). During the subsequent distributed fine-tuning phase, the server utilizes this aggregated decoder, $D_{\boldsymbol{\psi}^{\mathrm{global}}}$, while each client employs its locally adapted encoder. This setup aligns the compression module with heterogeneous client distributions without sacrificing the scalability of a unified server-side architecture.

\vspace{-5pt} \subsection{AutoEncoder-Compressed Distributed Fine-Tuning (DFT)} \vspace{-3pt}

During DFT, client $n$ retains its encoder $E_{\boldsymbol{\phi}_n}$ after CSA, while the server uses the aggregated decoder $D_{\boldsymbol{\psi}^{\mathrm{global}}}$. With both AE modules \textit{frozen}, the fine-tuning procedure is as follows:

\begin{enumerate}[leftmargin=*, itemsep=1pt, topsep=2pt]
    \item \textit{Client FP:} Client $n$ computes smashed data $(\mathbf{h}_{\mathrm{cls},n}, \mathbf{H}_n)$, compresses $\mathbf{H}_n$ as $\mathbf{Z}_n = E_{\boldsymbol{\phi}_n}(\mathbf{H}_n)$, and sends $(\mathbf{h}_{\mathrm{cls},n}, \mathbf{Z}_n)$ to the server.

    \item \textit{Server FP:} The server reconstructs $\widehat{\mathbf{H}}_n = D_{\boldsymbol{\psi}^{\mathrm{global}}}(\mathbf{Z}_n)$ and forwards $(\mathbf{h}_{\mathrm{cls},n}, \widehat{\mathbf{H}}_n)$ through $f_{\boldsymbol{\theta}_s}$ to obtain predictions $\hat{\mathbf{y}}_n$.

    \item \textit{Loss Aggregation:} Each client computes $\mathcal{L}_n$ using its local labels, and the server aggregates the received losses as $\mathcal{L}_g = \sum_{n \in \mathcal{N}} \mathcal{L}_n$.

    \item \textit{Server BP:} The server performs a single backpropagation pass, updates $f_{\boldsymbol{\theta}_s}$, and returns $\nabla_{\mathbf{Z}_n}\mathcal{L}_g$ and $\nabla_{\mathbf{h}_{\mathrm{cls},n}}\mathcal{L}_g$ to client $n$.

    \item \textit{Client BP:} Client $n$ backpropagates via the frozen encoder and updates $f_{\boldsymbol{\theta}_c}$.
\end{enumerate}

\noindent Since both $\mathbf{Z}_n$ and $\nabla_{\mathbf{Z}_n}\mathcal{L}_g$ operate in the reduced dimension $d_z \ll d$, AE-PSL reduces communication bi-directionally during DFT. Note that, while \method\ leverages SASL for its server-side efficiency, label privacy, and elimination of per-client sub-models (see Sec.~\ref{sec:background}), learnable compression inherently generalizes to other SL paradigms. The GA phase optimizes the AE for the general activation manifold of a pre-trained model, a structural property completely independent of the distributed training setup. Furthermore, although CSA is designed to aggregate decoders across parallel clients (as in SFL or SASL), it is equally applicable to standard sequential SL. In a sequential regime, the server can simply aggregate the client-adapted decoders at the conclusion of the local warm-up phase, establishing a unified global decoder that services all subsequent sequential fine-tuning rounds. Consequently, \method\ serves as a versatile communication-reduction solution applicable to any split-based fine-tuning workload.

\vspace{-5pt}\section{Experiments} 

\vspace{-5pt}\subsection{Datasets \& Models} \vspace{-3pt}
We evaluate \method\ on four public vision classification datasets, i.e., CIFAR-100 \cite{krizhevsky2009learning_cifar}, Food101 \cite{bossard2014food101}, SUN397 \cite{xiao2010sun397}, and FEMNIST \cite{DBLP:journals/corr/abs-1812-01097_femnist}. We partition client data using Dirichlet sampling~\cite{yurochkin2019bayesian-dirichlet} with $\alpha=0.5$ for all datasets except FEMNIST, which provides client-specific non-IID distributed splits. Unless stated otherwise, we use an ImageNet-pretrained ViT-B/32~\cite{dosovitskiy2021ViT,imagenet} partitioned at layer $s=5$ in our experiments. DFT is performed for $10$ epochs using LoRA~\cite{hu2022LoRA} with rank $r=16$ and $\alpha=32$, a global batch size of $125$, and Adam with learning rate $10^{-4}$. For GA, the AE is trained on ImageNet100~\cite{imagenet} for $30$ epochs. Due to space limitations, we provide the full experimental configuration in Appx.~\ref{app:exp_setup}, and analyze alternative split layers in Appx.~\ref{app:split_layer}. \par

\vspace{3pt} \noindent \textbf{AutoEncoder Architecture.} To balance reconstruction fidelity against client-side overhead, we evaluate two architecture types: token-wise \textit{MLP-based AEs} and spatial \textit{Convolution-based AEs}. Structurally native to transformers, \textit{MLP-based AEs} compress each token independently along the hidden dimension $d$, preserving sequence structure without spatial restructuring. Conversely \textit{Convolution-based AEs} reshape the 1D patch-token sequence into a 2D spatial grid prior to encoding to exploit localized spatial redundancies. This approach is inspired by existing SL learnable compression frameworks designed for convolutional networks \cite{eshratifar2019bottlenet,matsubara2022bottlefit,ayad2021slAE}. For MLP variants, we implement a symmetric configuration with the encoder denoted as $\ell\text{-layer-MLP-}(d\text{-}u_1\text{-}\cdots\text{-}d_z)$, where $\ell$ is the layer count, $u_i$ are intermediate widths, and $d_z$ is the compressed latent dimension. These layers incorporate Layer Normalization and \texttt{GELU} activations, which empirically outperformed \texttt{ReLU} (Appx.~\ref{app:activations}). Unless specified otherwise, we use a $2\text{-layer-MLP-}(d\text{-}d\text{-}d_z)$ architecture; detailed AE architectures analysis is in Sec.~\ref{exp:ae_arch}.

\vspace{-5pt}\subsection{Evaluation Setup} \vspace{-3pt}
\noindent \textbf{Scenarios.} Due to the split structure of PSL, DFT produces multiple client-side sub-models together with a single shared server-side sub-model, enabling different post-training evaluation modes. We consider two paradigms: \textit{global} evaluation, where client-side weights are aggregated into a single global client sub-model, combined with the shared server-side sub-model, and evaluated on a centralized test set with the AE bypassed; and \textit{local} evaluation, where each client is evaluated using its own client-side sub-model and local test data, with the AE active, and final performance is averaged across clients. Thus, \textit{global} evaluation isolates the effect of compression during DFT and tests whether the model remains compatible with the original uncompressed representation space at inference, while \textit{local} evaluation captures an end-to-end distributed setting where the communication bottleneck remains active during inference. Accordingly, we use FEMNIST for \textit{local} evaluation, as its user-partitioned data provides a separate test set for each client, enabling evaluation under client-specific distributions, while CIFAR-100, Food101, and SUN397 are used for \textit{global} evaluation as centralized image-classification benchmarks.

\vspace{3pt} \noindent \textbf{Baselines.} We compare \method{} against related SL heuristic-based communication-compression methods, covering batch-wise compression (ADC~\cite{alvetreti2025attention-dc}, C3-SL~\cite{hsieh2022c3-sl}) and top-$k$ sparsification (Rand-Top-$K$~\cite{zheng2023rand-topk}). For a controlled comparison, we integrate all baselines into the SASL framework, although the original methods were evaluated under traditional SL or SFL settings; for each baseline, the \cls~token is left uncompressed when doing so improves accuracy.

\vspace{3pt} \noindent \textbf{Metrics.} We report downstream task accuracy, client-side computational overhead measured in GFLOPs, and communication reduction $R$, defined relative to uncompressed DFT. For fair comparison, we evaluate each method $3$ reduction levels: \textit{Low} ($R\approx5\times$), \textit{Mid} ($R\approx10\times$), and \textit{High} ($R\approx20\times$), spanning from light compression to a high compression ratio where baselines experience significant degradation (see Sec. \ref{ssec:results}). Note that $R$ is upper-bounded by a hard ceiling, as the uncompressed \cls\ token requires a fixed per-iteration communication budget. All results are averaged over $3$ random seeds.

\begin{table}[!t]
  \centering \footnotesize
  \setlength{\tabcolsep}{2.4pt}
  \renewcommand{\arraystretch}{1.06}
  \caption{Performance under \textbf{\textit{global}} and \textbf{\textit{local}} evaluation using ViT-B/32~\cite{dosovitskiy2021ViT} across datasets for various $N$ and $R$. We report average accuracy across $3$ seeds, whereas \textit{deltas} denote relative performance to the \textit{no compression} baseline ($R$=$1$) for the same settings. Results with standard deviations are reported in Table~\ref{tab:std_main_results}.}
  \label{tab:main_results}

  \resizebox{\linewidth}{!}{%
  \begin{tabular}{c l l ccc ccc}
    \toprule
    & \multirow{2}{*}{\textbf{Dataset}} & \multirow{2}{*}{\textbf{Method}}
    & \multicolumn{3}{c}{$N = 5$} & \multicolumn{3}{c}{$N = 25$} \\
    \cmidrule(lr){4-6} \cmidrule(lr){7-9}
    & & & Low ($R\approx5$) & Mid ($R\approx10$) & High ($R\approx 20$)
      & Low ($R\approx5$) & Mid ($R\approx10$) & High ($R\approx 20$) \\
    \midrule

    \multirow{15}{*}{\rotatebox[origin=c]{90}{\textit{Global Eval.}}}
    & \multirow{5}{*}{\textbf{CIFAR100}}
    & No Compr. ($R=1$) & \multicolumn{3}{c}{$82.7$} & \multicolumn{3}{c}{$82.0$} \\
    \cmidrule(lr){3-9}
    & & C3-SL      & $66.5$ {\scriptsize\textcolor{dblue}{$(-16.2)$}} & $50.9$ {\scriptsize\textcolor{dblue}{$(-31.8)$}} & $37.3$ {\scriptsize\textcolor{dblue}{$(-45.4)$}} & $60.0$ {\scriptsize\textcolor{dblue}{$(-22.0)$}} & $60.3$ {\scriptsize\textcolor{dblue}{$(-21.7)$}} & $59.9$ {\scriptsize\textcolor{dblue}{$(-22.1)$}} \\
    & & Rand-Top-K & $79.9$ {\scriptsize\textcolor{dblue}{$(-2.8)$}}  & $76.2$ {\scriptsize\textcolor{dblue}{$(-6.5)$}}  & $61.5$ {\scriptsize\textcolor{dblue}{$(-21.2)$}} & $78.8$ {\scriptsize\textcolor{dblue}{$(-3.2)$}}  & $75.4$ {\scriptsize\textcolor{dblue}{$(-6.6)$}}  & $63.3$ {\scriptsize\textcolor{dblue}{$(-18.7)$}} \\
    & & ADC        & $78.0$ {\scriptsize\textcolor{dblue}{$(-4.7)$}}  & $68.4$ {\scriptsize\textcolor{dblue}{$(-14.3)$}} & $53.7$ {\scriptsize\textcolor{dblue}{$(-29.0)$}} & $71.7$ {\scriptsize\textcolor{dblue}{$(-10.3)$}} & $68.0$ {\scriptsize\textcolor{dblue}{$(-14.0)$}} & $63.4$ {\scriptsize\textcolor{dblue}{$(-18.6)$}} \\
    \cmidrule(lr){3-9}
    & & \method{} (Ours) & $\mathbf{83.0}$ {\scriptsize\textcolor{dblue}{$\mathbf{(+0.3)}$}} & $\mathbf{82.7}$ {\scriptsize\textcolor{dblue}{$\mathbf{(+0.0)}$}} & $\mathbf{81.4}$ {\scriptsize\textcolor{dblue}{$\mathbf{(-1.3)}$}} & $\mathbf{82.7}$ {\scriptsize\textcolor{dblue}{$\mathbf{(+0.7)}$}} & $\mathbf{82.7}$ {\scriptsize\textcolor{dblue}{$\mathbf{(+0.7)}$}} & $\mathbf{81.1}$ {\scriptsize\textcolor{dblue}{$\mathbf{(-0.9)}$}} \\
    \cmidrule(lr){2-9}

    & \multirow{5}{*}{\textbf{Food101}}
    & No Compr. ($R=1$) & \multicolumn{3}{c}{$69.1$} & \multicolumn{3}{c}{$67.5$} \\
    \cmidrule(lr){3-9}
    & & C3-SL      & $37.0$ {\scriptsize\textcolor{dblue}{$(-32.1)$}} & $40.3$ {\scriptsize\textcolor{dblue}{$(-28.8)$}} & $32.1$ {\scriptsize\textcolor{dblue}{$(-37.0)$}} & $31.3$ {\scriptsize\textcolor{dblue}{$(-36.2)$}} & $31.3$ {\scriptsize\textcolor{dblue}{$(-36.2)$}} & $31.3$ {\scriptsize\textcolor{dblue}{$(-36.2)$}} \\
    & & Rand-Top-K & $66.5$ {\scriptsize\textcolor{dblue}{$(-2.6)$}}  & $60.2$ {\scriptsize\textcolor{dblue}{$(-8.9)$}}  & $35.7$ {\scriptsize\textcolor{dblue}{$(-33.4)$}} & $65.1$ {\scriptsize\textcolor{dblue}{$(-2.4)$}}  & $60.3$ {\scriptsize\textcolor{dblue}{$(-7.2)$}}  & $43.1$ {\scriptsize\textcolor{dblue}{$(-24.4)$}} \\
    & & ADC        & $63.3$ {\scriptsize\textcolor{dblue}{$(-5.8)$}}  & $54.6$ {\scriptsize\textcolor{dblue}{$(-14.5)$}} & $44.0$ {\scriptsize\textcolor{dblue}{$(-25.1)$}} & $56.2$ {\scriptsize\textcolor{dblue}{$(-11.3)$}} & $51.8$ {\scriptsize\textcolor{dblue}{$(-15.7)$}} & $47.1$ {\scriptsize\textcolor{dblue}{$(-20.4)$}} \\
    \cmidrule(lr){3-9}
    & & \method{} (Ours) & $\mathbf{70.3}$ {\scriptsize\textcolor{dblue}{$\mathbf{(+1.2)}$}} & $\mathbf{69.6}$ {\scriptsize\textcolor{dblue}{$\mathbf{(+0.5)}$}} & $\mathbf{65.3}$ {\scriptsize\textcolor{dblue}{$\mathbf{(-3.8)}$}} & $\mathbf{68.7}$ {\scriptsize\textcolor{dblue}{$\mathbf{(+1.2)}$}} & $\mathbf{68.1}$ {\scriptsize\textcolor{dblue}{$\mathbf{(+0.6)}$}} & $\mathbf{64.3}$ {\scriptsize\textcolor{dblue}{$\mathbf{(-3.2)}$}} \\
    \cmidrule(lr){2-9}

    & \multirow{5}{*}{\textbf{SUN397}}
    & No Compr. ($R=1$) & \multicolumn{3}{c}{$67.0$} & \multicolumn{3}{c}{$64.5$} \\
    \cmidrule(lr){3-9}
    & & C3-SL      & $48.0$ {\scriptsize\textcolor{dblue}{$(-19.0)$}} & $41.3$ {\scriptsize\textcolor{dblue}{$(-25.7)$}} & $31.6$ {\scriptsize\textcolor{dblue}{$(-35.4)$}} & $41.5$ {\scriptsize\textcolor{dblue}{$(-23.0)$}} & $41.5$ {\scriptsize\textcolor{dblue}{$(-23.0)$}} & $41.5$ {\scriptsize\textcolor{dblue}{$(-23.0)$}} \\
    & & Rand-Top-K & $66.3$ {\scriptsize\textcolor{dblue}{$(-0.7)$}}  & $62.9$ {\scriptsize\textcolor{dblue}{$(-4.1)$}}  & $48.5$ {\scriptsize\textcolor{dblue}{$(-18.5)$}} & $64.6$ {\scriptsize\textcolor{dblue}{$(+0.1)$}} & $61.2$ {\scriptsize\textcolor{dblue}{$(-3.3)$}} & $50.0$ {\scriptsize\textcolor{dblue}{$(-14.5)$}} \\
    & & ADC        & $62.2$ {\scriptsize\textcolor{dblue}{$(-4.8)$}}  & $52.9$ {\scriptsize\textcolor{dblue}{$(-14.1)$}} & $39.0$ {\scriptsize\textcolor{dblue}{$(-28.0)$}} & $55.5$ {\scriptsize\textcolor{dblue}{$(-9.0)$}}  & $52.1$ {\scriptsize\textcolor{dblue}{$(-12.4)$}} & $47.5$ {\scriptsize\textcolor{dblue}{$(-17.0)$}} \\
    \cmidrule(lr){3-9}
    & & \method{} (Ours) & $\mathbf{67.4}$ {\scriptsize\textcolor{dblue}{$\mathbf{(+0.4)}$}} & $\mathbf{66.7}$ {\scriptsize\textcolor{dblue}{$\mathbf{(-0.3)}$}} & $\mathbf{64.8}$ {\scriptsize\textcolor{dblue}{$\mathbf{(-2.2)}$}} & $\mathbf{65.2}$ {\scriptsize\textcolor{dblue}{$\mathbf{(+0.7)}$}} & $\mathbf{65.0}$ {\scriptsize\textcolor{dblue}{$\mathbf{(+0.5)}$}} & $\mathbf{63.8}$ {\scriptsize\textcolor{dblue}{$\mathbf{(-0.7)}$}} \\

    \specialrule{0.8pt}{3pt}{3pt}

    \multirow{5}{*}{\rotatebox[origin=c]{90}{\textit{Local Eval.}}}
    & \multirow{5}{*}{\textbf{FEMNIST}}
    & No Compr. ($R=1$) & \multicolumn{3}{c}{$87.9$} & \multicolumn{3}{c}{$85.8$} \\
    \cmidrule(lr){3-9}
    & & C3-SL      & $85.2$ {\scriptsize\textcolor{dblue}{$(-2.7)$}} & $83.5$ {\scriptsize\textcolor{dblue}{$(-4.4)$}} & $81.4$ {\scriptsize\textcolor{dblue}{$(-6.5)$}} & $80.7$ {\scriptsize\textcolor{dblue}{$(-5.1)$}} & $39.8$ {\scriptsize\textcolor{dblue}{$(-46.0)$}} & $6.9$ {\scriptsize\textcolor{dblue}{$(-78.9)$}} \\
    & & Rand-Top-K & $87.6$ {\scriptsize\textcolor{dblue}{$(-0.3)$}}  & $87.2$ {\scriptsize\textcolor{dblue}{$(-0.7)$}}  & $86.4$ {\scriptsize\textcolor{dblue}{$(-1.5)$}} & $84.6$ {\scriptsize\textcolor{dblue}{$(-1.2)$}}  & $83.8$ {\scriptsize\textcolor{dblue}{$(-2.0)$}}  & $82.2$ {\scriptsize\textcolor{dblue}{$(-3.6)$}} \\
    & & ADC        & $86.6$ {\scriptsize\textcolor{dblue}{$(-1.3)$}}  & $81.1$ {\scriptsize\textcolor{dblue}{$(-6.8)$}} & $71.8$ {\scriptsize\textcolor{dblue}{$(-16.1)$}} & $82.4$ {\scriptsize\textcolor{dblue}{$(-3.4)$}} & $78.1$ {\scriptsize\textcolor{dblue}{$(-7.7)$}} & $73.8$ {\scriptsize\textcolor{dblue}{$(-12.0)$}} \\
    \cmidrule(lr){3-9}
    & & \method{} (Ours) & $\mathbf{88.1}$ {\scriptsize\textcolor{dblue}{$\mathbf{(+0.2)}$}} & $\mathbf{88.0}$ {\scriptsize\textcolor{dblue}{$\mathbf{(+0.1)}$}} & $\mathbf{87.9}$ {\scriptsize\textcolor{dblue}{$\mathbf{(+0.0)}$}} & $\mathbf{85.9}$ {\scriptsize\textcolor{dblue}{$\mathbf{(+0.1)}$}} & $\mathbf{85.5}$ {\scriptsize\textcolor{dblue}{$\mathbf{(-0.3)}$}} & $\mathbf{84.6}$ {\scriptsize\textcolor{dblue}{$\mathbf{(-1.2)}$}} \\
    \bottomrule
  \end{tabular}%
  }
\end{table}

\subsection{Results \& Discussion} \label{ssec:results}

\subsubsection{Accuracy--Communication Trade-off} \hfill \\
We first examine the central accuracy--communication trade-off: whether each method preserves downstream performance as communication is reduced. We evaluate all methods under \textit{global} and \textit{local} evaluation using ViT-B/32, with $N\in\{5,25\}$ clients and $R\in\{5,10,20\}$. Table~\ref{tab:main_results} shows that \method\ remains close to the \textit{no compression} baseline across datasets, client counts, and reduction levels. At low and mid reductions ($R\approx5$ and $R\approx10$), \method\ often matches or exceeds the uncompressed baseline, suggesting that the AE bottleneck can regularize intermediate representations by suppressing task-irrelevant noise while preserving semantic features. Even at high reduction ($R\approx20$), the largest drop is $3.8$ percentage points (pp.). \par

Compared to the baselines, \method~achieves the best accuracy–communication trade-off, attaining the highest accuracy across both \textit{global} and \textit{local} evaluation scenarios for all compression levels. Under \textit{global} evaluation, where the AE is bypassed at inference, \method~consistently preserves compatibility with the uncompressed intermediate representation space, while heuristic baselines degrade sharply as $R$ increases, especially on Food101 and SUN397. Under \textit{local} evaluation on FEMNIST, performance of baselines is less affected at $R\approx5$, yet the gap widens at higher compression levels. Specifically, \method~remains within $1.21$ pp. of the uncompressed baseline at $R\approx20$, whereas ADC and C3-SL degrade substantially, with C3-SL collapsing for $N=25$. Since the global batch is fixed at $125$, increasing $N$ from $5$ to $25$ reduces each clients' mini-batch size; yet, these batch-wise compression methods do not systematically worsen, suggesting that their degradation is driven more by static compression than merely by batch size. Overall, these results highlight that \textit{learnable AE-based compression provides a more robust accuracy–communication trade-off than heuristic compression} across both evaluation modes: it preserves a deployable global model when the AE is bypassed after DFT, while maintaining client-specific accuracy when the communication bottleneck remains active at inference.


\begin{figure}[!t]
    \centering \small
    \begin{subfigure}{\linewidth}
        \centering \small        
        \includegraphics[width=\linewidth]{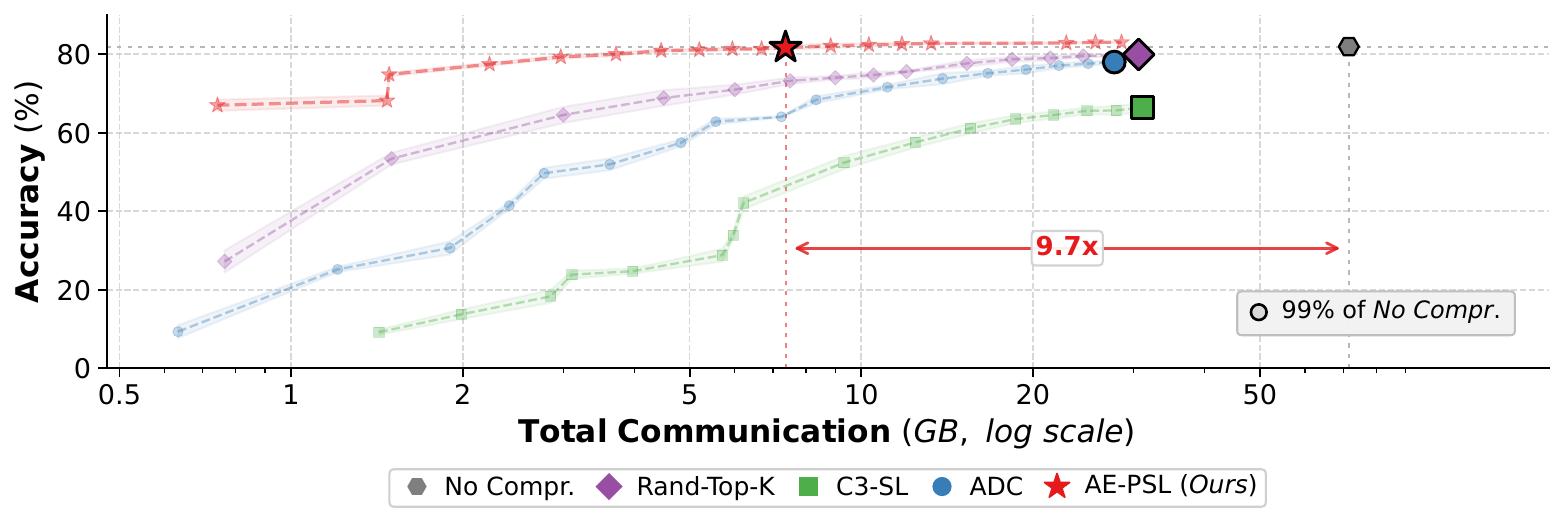}
        \caption{$N=5$} \label{fig:comm-cifar100-n5}
    \end{subfigure}
    \begin{subfigure}{\linewidth}
        \centering \small 
        \includegraphics[width=\linewidth]{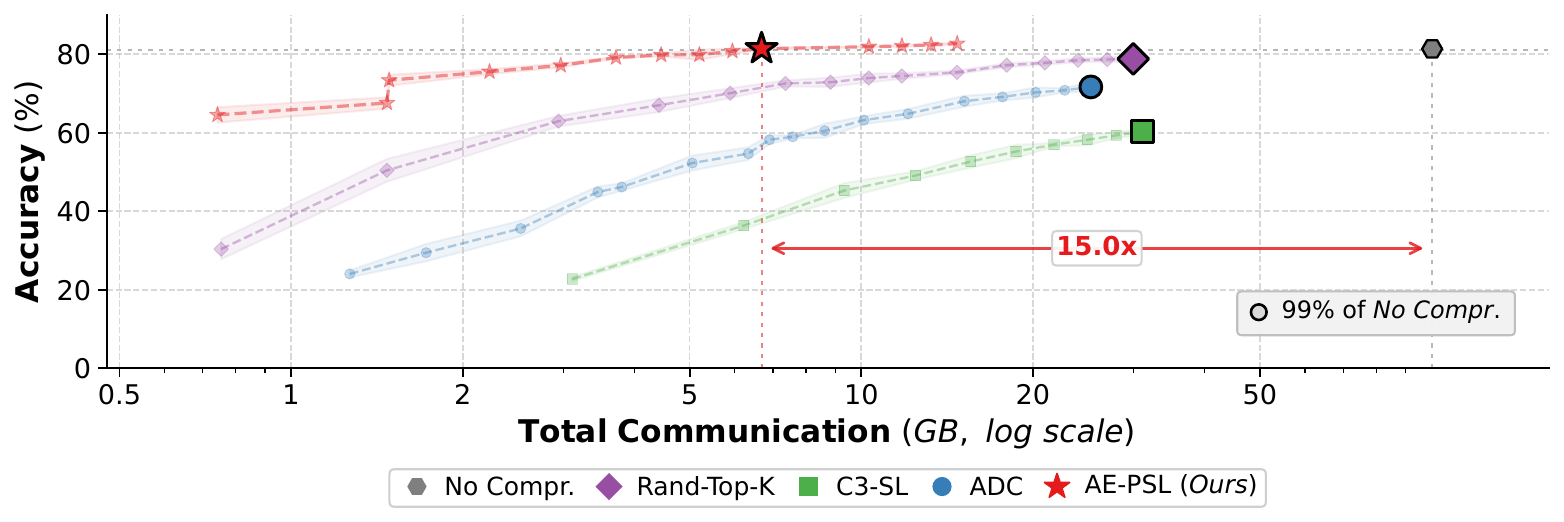}
        \caption{$N=25$} \label{fig:comm-cifar100-n25}
    \end{subfigure}
    \caption{Accuracy--communication trade-off using ViT-B/32 on CIFAR-100. We report average accuracy across $3$ seeds for different communication volumes (GB); markers indicate \textit{closest} (within $99\%$) of \textit{no compression} DFT.}
    \label{fig:comm}
\end{figure}

\vspace{3pt} \noindent \textbf{\textit{Fixed Communication Budget}.} We further evaluate accuracy as a function of total communication volume across clients. As shown in Fig.~\ref{fig:comm} for CIFAR-100, \method~reaches near-peak accuracy substantially earlier than all baselines for both $N=5$ and $N=25$. Specifically, across all datasets and client settings, \method~reaches no-compression DFT accuracy with $10.2\times$ less communication on average; at the same communication budget, the strongest evaluated baseline, Rand-Top-K, trails \method~by $5.3$ pp. on average. Furthermore, to reach  within 99\% of no-compression DFT \method~needs $12.4\times$ less communication budget, with Rand-Top-K trailing by $13.9$ pp. on average. This indicates that \method~provides more communication-efficient convergence, as \textit{its learned, feature-distribution-aware AE module preserves more task-relevant intermediate information with fewer transmitted bytes than heuristic compression modules.} Per-dataset results are reported in Appx.~\ref{app:extra_baselines}.

\begin{figure}[!t]
    \centering \small
    \includegraphics[width=\linewidth]{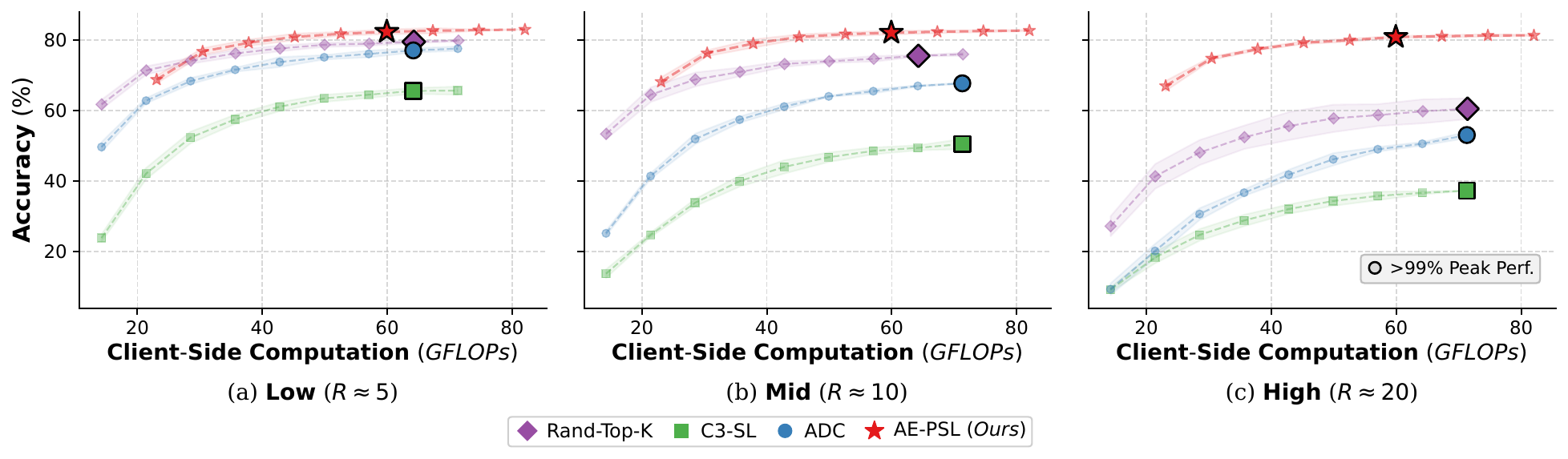}
    \caption{Accuracy--client-side compute trade-off using ViT-B/32 for $N=5$ and $R\in\{5,10,20\}$ on CIFAR-100. We report average accuracy across $3$ seeds for cumulative client-side computation (GFLOPs) across compression levels; markers indicate when a method reaches $>99\%$ of its peak performance.}
    \label{fig:comp}
\end{figure}


\begin{figure}[!t]
    \centering \small
    \includegraphics[width=\linewidth]{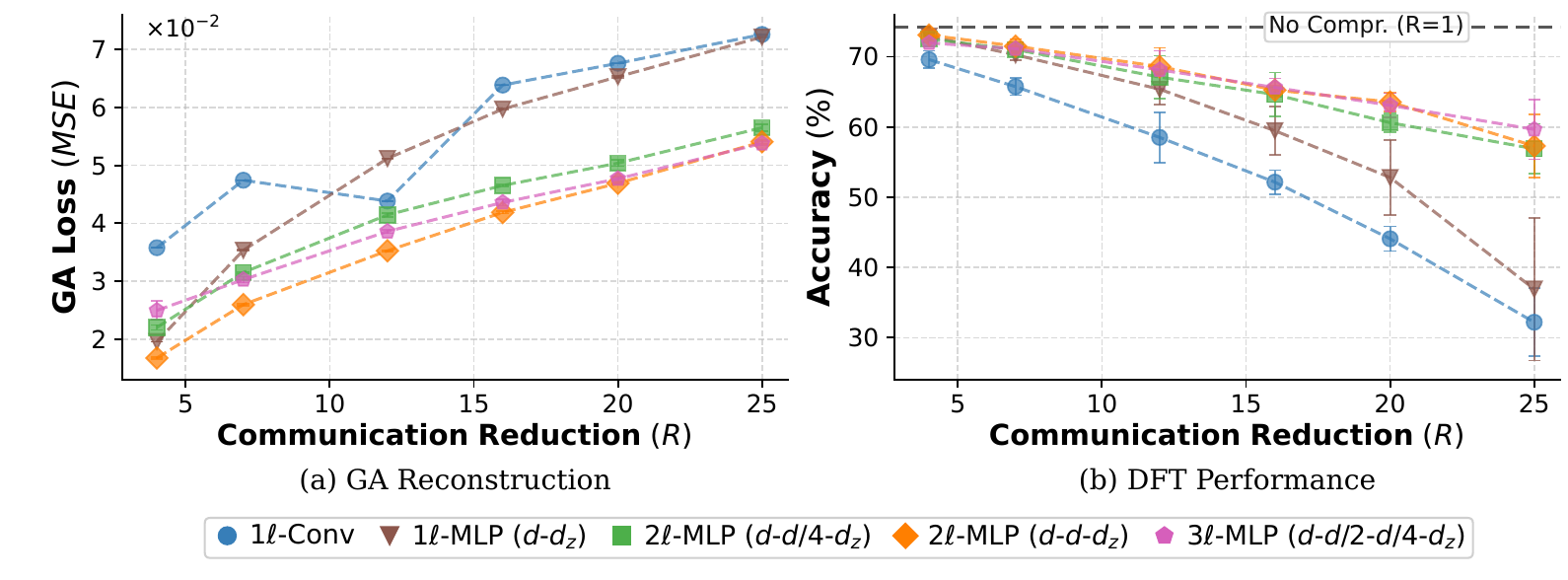}
    \caption{Effect of AE architecture in~\method~across communication reduction levels ($R$). Experiments are conducted with ViT-B/32~\cite{dosovitskiy2021ViT} on Food101 with $N=1$. We report average (a) MSE reconstruction loss during GA, and (b) downstream DFT accuracy over $3$ seeds.}
    \label{fig:ae_arch}
\end{figure}

\vspace{-5pt}
\subsubsection{Client-side Computational Overhead} \hfill \\
We next evaluate whether the accuracy gains of \method~come at excessive client-side computation cost. For this, we measure cumulative client-side GFLOPs during DFT, including the CSA warm-up cost of~\method. As shown in Fig.~\ref{fig:comp} for CIFAR-100 with $N=5$, \method~consistently reaches a higher accuracy regime than the baselines across all compression levels. To reach $>99\%$ of peak performance, the baselines require on average $1.07\times$, $1.15\times$, and $1.19\times$ more GFLOPs than \method~at low, mid, and high reduction levels, respectively. This shows that the lightweight encoder and CSA warm-up overhead are outweighed by improved compute efficiency during DFT: as $R$ increases, \textit{\method~reaches near-peak performance with fewer client-side GFLOPs by using learnable feature-preserving compression, whereas heuristic baselines require more computation and still converge to lower peak accuracies.} Results for the remaining datasets are reported in Appx.~\ref{app:extra_baselines}, where the same pattern is observed.

\vspace{-5pt} 

\vspace{-5pt}
\subsubsection{Impact of AutoEncoder Architecture} \label{exp:ae_arch}  \hfill \\ 
We analyze the effect of AE architecture on reconstruction fidelity and downstream accuracy. For each design, we first perform General Alignment (GA; Sec.\ref{sec:GA}) and measure the resulting MSE reconstruction loss, then integrate the aligned AE into DFT on Food101 to assess downstream accuracy. To isolate architectural effects from client heterogeneity and decoder aggregation, we use a single-client setting ($N=1$). \par

As shown in Fig.\ref{fig:ae_arch}, convolutional AEs are less robust across compression levels, suggesting that reshaping transformer patch-token sequences into 2D grids often leads to lower downstream accuracy and higher reconstruction loss. For MLP-based AEs, depth primarily matters under stronger compression: the 3-layer MLP achieves the highest accuracy at extreme reduction levels (e.g., $R\approx25$), with limited gains at lower compression. Conversely, 1-layer architectures remain close to the no-compression setting ($R=1$) for $R<10$, yet their accuracy drops sharply as $R$ increases. The results also reveal a clear inverse relationship between GA reconstruction loss and downstream DFT accuracy, indicating that preserving split-layer features is critical for effective fine-tuning. Based on this trade-off, we select $2\text{-layer-MLP-}(d\text{-}d\text{-}d_z)$ as the default architecture, as it achieves strong reconstruction and downstream accuracy while increasing client-side parameters by only $1.6\%$. We further analyze the accuracy--parameter trade-off in Appx.~\ref{sec:app_ae_param}.

\begin{table}[!b]
  \centering \scriptsize
  \setlength{\tabcolsep}{3.2pt}
  \renewcommand{\arraystretch}{1.08}
  \caption{Effect of \method's core components: General Alignment (GA), Client-Side Alignment (CSA), and AE freezing during DFT (FZ). Experiments conducted with ViT-B/32~\cite{dosovitskiy2021ViT} under \textit{global} and \textit{local} evaluation for $R=15$ with $N\in\{5,25\}$. We report average accuracy across $3$ seeds; deltas denote relative performance to \textit{no compression} baselines ($R$=$1$) for the same settings.}
  \label{tab:srq2_ablation}
  \resizebox{\linewidth}{!}{%
  \begin{tabular}{l l ccc c c}
    \toprule
    \multirow{2}{*}{\textbf{$N$}} & \multirow{2}{*}{\textbf{Method}}
    & \multicolumn{3}{c}{\textit{Global evaluation}}
    & \multicolumn{1}{c}{\textit{Local evaluation}}
    & \multirow{2}{*}{\textbf{R}} \\
    \cmidrule(lr){3-5} \cmidrule(lr){6-6}
    & & \textbf{CIFAR100} & \textbf{Food101} & \textbf{SUN397} & \textbf{FEMNIST} & \\
    \midrule
    \multirow{5}{*}{$N=5$}
    & No Compression & $82.7 \pm 0.4$ & $69.1 \pm 2.1$ & $67.0 \pm 1.2$ & $87.6 \pm 0.7$ & --- \\
    \cmidrule(lr){2-7}
    & AE & \abres{1.4}{0.3}{-81.3} & \abres{1.1}{0.2}{-68.0} & \abres{0.3}{0.2}{-66.7} & \abres{86.9}{2.0}{-0.7} & $15\times$ \\
    & AE \!+\! GA & \abres{75.2}{0.4}{-7.5} & \abres{63.2}{2.8}{-5.9} & \abres{62.8}{0.4}{-4.2} & \abres{87.1}{1.1}{-0.5} & $15\times$ \\
    & AE \!+\! GA \!+\! FZ & \abres{79.6}{0.9}{-3.1} & \abres{67.2}{3.2}{-1.9} & \abres{64.9}{0.6}{-2.1} & \abresbest{88.0}{0.6}{+0.4} & $15\times$ \\
    & AE \!+\! GA \!+\! CSA \!+\! FZ (\method) & \abresbest{82.2}{0.7}{-0.5} & \abresbest{67.6}{2.5}{-1.5} & \abresbest{65.6}{0.7}{-1.4} & \abres{88.0}{0.7}{+0.4} & $15\times$ \\
    \midrule
    \multirow{5}{*}{$N=25$}
    & No Compression & $82.0 \pm 0.1$ & $67.5 \pm 0.1$ & $64.5 \pm 2.7$ & $85.4 \pm 1.3$ & --- \\
    \cmidrule(lr){2-7}
    & AE & \abres{1.1}{0.1}{-80.9} & \abres{1.3}{0.2}{-66.2} & \abres{0.5}{0.6}{-64.0} & \abres{84.3}{1.8}{-1.1} & $15\times$ \\
    & AE \!+\! GA & \abres{74.2}{2.3}{-7.8} & \abres{61.8}{0.7}{-5.7} & \abres{61.2}{2.1}{-3.3} & \abresbest{85.5}{1.1}{+0.1} & $15\times$ \\
    & AE \!+\! GA \!+\! FZ & \abres{79.0}{1.5}{-3.0} & \abres{65.7}{0.4}{-1.8} & \abres{63.4}{1.1}{-1.1} & \abres{84.8}{0.9}{-0.6} & $15\times$ \\
    & AE \!+\! GA \!+\! CSA \!+\! FZ (\method) & \abresbest{82.2}{0.5}{+0.2} & \abresbest{66.7}{1.3}{-0.8} & \abresbest{64.8}{1.3}{+0.3} & \abres{84.1}{0.8}{-1.3} & $15\times$ \\
    \bottomrule
  \end{tabular}%
  }
\end{table}

\vspace{-5pt} \subsubsection{\method~Components Analysis}\label{sec:training_integration_ablation}\hfill\\ 
We now ablate the core components of \method: General Alignment (GA), Client-Side Alignment (CSA), and AE freezing during DFT (FZ). As shown in Table~\ref{tab:srq2_ablation}, directly inserting a randomly initialized AE causes severe collapse under \textit{global} evaluation.
Alternatively, we note that each component addresses a different source of instability. GA substantially restores global accuracy by aligning the AE with the pre-trained feature manifold before DFT, while AE freezing during DFT further improves performance by keeping the compression mapping stable during fine-tuning, preventing reconstruction quality drift. Finally, CSA adapts the encoder to client-specific feature distributions while retaining a single aggregated decoder, further closing the gap to no-compression settings. Together, these components make learnable compression compatible with off-the-shelf pre-trained models, approaching near-uncompressed performance while achieving substantial communication reduction. Appx.~\ref{app:local_eval} expands on \textit{local} evaluation results. Appx.~\ref{app:training_integration} further analyzes components by visualizing AE reconstruction fidelity throughout DFT. Appx.~\ref{app:alternative_alignment} details alternative AE alignment strategies.


\vspace{-5pt}

\vspace{-5pt}\section{Conclusion} \vspace{-3pt}
We present \method\, a framework designed for communication-efficient distributed fine-tuning of large-scale FMs on resource-constrained edge devices. By integrating a lightweight, \textit{learnable AE} into PSL, \method\ effectively compresses intermediate feature representations, substantially reducing the volume of data transmitted between clients and the server. 
Whereas existing learnable compression methods showed promising results, these are inherently incompatible with pre-trained models causing catastrophic forgetting due to feature misalignment. To mitigate this, we introduce 2-stage alignment. This process---comprising GA and CSA---tailors learnable compression to each clients' feature distribution prior to fine-tuning, without centralizing private data. We evaluate \method\ across $4$ vision datasets, showing that feature-aligned learnable compression provides a more robust accuracy--communication trade-off than heuristic SL compression, achieving 10.2× communication reduction without accuracy degradation. 
Furthermore, \method\ accelerates distributed convergence, reaching 99\% of peak accuracy using 12.4× less total communication volume than heuristic baseline methods. Crucially, this communication efficiency does not rely on excessive local computation; \method\ requires up to 1.14× fewer client-side GFLOPs to achieve near-peak performance compared to heuristic alternatives. By introducing communication overhead reduction while preserving downstream task accuracy, \method\ establishes a scalable and communication-efficient pathway for collaborative pre-trained model fine-tuning in resource-constrained edge environments.

\vspace{3pt}\noindent \textbf{Limitations \& Future Work.} 
 Although conceptually our two-stage alignment protocol for learnable compression generalizes to other modalities, its performance across non-vision domains remains unevaluated. Additionally, \method\ currently relies on transmitting the \cls\ token uncompressed to guarantee downstream accuracy. While highly effective for image classification, this structural dependency limits the framework's broader architectural scope. Recognizing that not all tokens hold equal semantic importance, future work will explore token-adaptive compression via learnable importance scores. This approach aims to eliminate the dependency on uncompressed global tokens, paving the way for a more versatile, model-agnostic edge fine-tuning framework.

\section*{Acknowledgments}
This work was supported by the AIMS5.0 project under grant agreement no.~101112089.




\bibliographystyle{splncs04}
\bibliography{references}

\newpage
\appendix
\setcounter{section}{0}
\renewcommand{\thesection}{\Alph{section}}

\section{Appendix}


\subsection{Detailed Experimental Setup} \label{app:exp_setup}
We evaluate \method\ on public vision classification datasets. CIFAR-100 \cite{krizhevsky2009learning_cifar}, Food101 \cite{bossard2014food101} and SUN397 \cite{xiao2010sun397} do not provide per-client test sets, and are used for \textit{global} evaluation. If a dataset is not pre-split, 20\% is left out for evaluation. We partition client data using Dirichlet sampling~\cite{yurochkin2019bayesian-dirichlet} with $\alpha=0.5$. Under \textit{local evaluation}, we use  FEMNIST \cite{DBLP:journals/corr/abs-1812-01097_femnist}, a dataset specifically collected for distributed learning settings, providing per-user datasets. For \textit{local evaluation} on FEMNIST, the dataset is inherently partitioned by \texttt{writer\_id}, with the total pool of writers distributed evenly and in a non-overlapping manner across the participating clients. To manage computational constraints, we subsample 10\% of the data samples from each assigned writer before applying localized train and test splits.

\vspace{3pt}\noindent \textbf{Training Parameters.}
DFT is performed for $10$ epochs using LoRA~\cite{hu2022LoRA} with rank $r=16$ and $\alpha=32$, optimized via Cross Entropy loss using Adam \cite{kingma2015adam} with learning rate $10^{-4}$, which is scheduled using Cosine Annealing, bringing the rate down to $5\times 10^{-5}$ over the duration of DFT. The global batch size is fixed at $125$, hence client batch size are $25, 5$ for $N=5,25$ respectively. For GA, the AE is trained on ImageNet100~\cite{imagenet} for $30$ epochs using Adam with with learning rate $10^{-4}$ with a Cosine Annealing scheduler annealing the rate down to $5\times 10^{-5}$ over $30$ epochs. Batch size is fixed at $512$. In the case of CSA, the client-side model partition remains entirely frozen, except for the AE. The AE module is optimized using Adam, with a higher learning rate of $10^{-3}$ over a single epoch. Client batch size are equal to the DFT client batch size.

\vspace{3pt}\noindent \textbf{Baselines.} In order to ensure fair comparison, we fix 3 communication reduction levels ($R$) for our method and all baselines. This accounts for optional compression of the \cls\ token, which is disabled if it improves performance for a given baseline. Specific parameters setting for each method are given in Table \ref{tab:baseline_parameters}. Each method performs DFT 3 times. All reported values are averaged over these 3 runs.

\begin{table}[!h]
  \centering \scriptsize
  \setlength{\tabcolsep}{4.2pt}
  \renewcommand{\arraystretch}{1.12}
  \caption{Compression-parameter settings used for each method across communication-reduction regimes. CLS tokens are left uncompressed unless indicated otherwise.}
  \label{tab:baseline_parameters}
  \resizebox{\linewidth}{!}{%
  \begin{tabular}{l l c c c c}
    \toprule
    \multirow{2}{*}{\textbf{Method}} &
    \multirow{2}{*}{\textbf{Parameter}} &
    \multicolumn{3}{c}{\textbf{Communication Reduction ($R$)}} &
    \multirow{2}{*}{\textbf{CLS Compr.}} \\
    \cmidrule(lr){3-5}
    & &
    \textbf{Low ($\sim5\times$)} &
    \textbf{Mid ($\sim10\times$)} &
    \textbf{High ($\sim20\times$)} &
    \\
    \midrule
    \textbf{AE-PSL (Ours)}
    & Latent dimension
    & $140$ & $64$ & $24$ & No \\

    Rand-Top-K~\cite{zheng2023rand-topk}
    & Sparsity
    & $0.15$ & $0.065$ & $0.025$ & No \\

    C3-SL~\cite{hsieh2022c3-sl}
    & Batch-wise CR
    & $5$ & $12$ & $34$ & No \\

    ADC~\cite{alvetreti2025attention-dc}$^\ast$
    & Batch/token CR
    & $1/\sqrt{5}$ & $1/\sqrt{10}$ & $1/\sqrt{20}$ & Yes \\
    \bottomrule
  \end{tabular}%
  }
\end{table}


\newpage
\subsection{Supplementary Performance Comparison with Baselines} \label{app:extra_baselines}
To evaluate performance of~\method~compared to the established baselines, we measure top-1 accuracy at varying degrees of communication reduction. Table \ref{tab:std_main_results} contains results with standard deviations included. Our proposal achieves the highest performance in terms of accuracy across all baselines. 

\begin{table}[]
  \centering \footnotesize
  \setlength{\tabcolsep}{2.4pt}
  \renewcommand{\arraystretch}{1.06}
  \caption{Performance under \textbf{\textit{global}} and \textbf{\textit{local}} evaluation using ViT-B/32~\cite{dosovitskiy2021ViT} across datasets for $N\in\{5,25\}$ and $R\in\{5,10,20\}$. We report average accuracy across $3$ seeds including standard deviations, whereas \textit{deltas} denote relative performance to \textit{no compression} baseline ($R$=$1$) for the same settings.} \label{tab:std_main_results}

  \resizebox{\linewidth}{!}{%
  \begin{tabular}{c l l ccc ccc}
    \toprule
    & \multirow{2}{*}{\textbf{Dataset}} & \multirow{2}{*}{\textbf{Method}}
    & \multicolumn{3}{c}{$N = 5$} & \multicolumn{3}{c}{$N = 25$} \\
    \cmidrule(lr){4-6} \cmidrule(lr){7-9}
    & & & Low ($R\approx 5$) & Mid ($R\approx 10$) & High ($R\approx 20$)
      & Low ($R\approx 5$) & Mid ($R\approx 10$) & High ($R\approx 20$) \\
    \midrule

    \multirow{18}{*}{\rotatebox[origin=c]{90}{\textit{Global Eval.}}}
    & \multirow{5}{*}{\textbf{CIFAR100}}
    & No Compr. ($R=1$) & \multicolumn{3}{c}{$82.7 \pm 0.2$} & \multicolumn{3}{c}{$82.0 \pm 0.1$} \\
    \cmidrule(lr){3-9}
    & & C3-SL      & $66.5 \pm 1.3$ {\scriptsize\textcolor{dblue}{$(-16.2)$}} & $50.9 \pm 1.2$ {\scriptsize\textcolor{dblue}{$(-31.8)$}} & $37.3 \pm 0.2$ {\scriptsize\textcolor{dblue}{$(-45.4)$}} & $60.0 \pm 0.5$ {\scriptsize\textcolor{dblue}{$(-22.0)$}} & $60.3 \pm 1.0$ {\scriptsize\textcolor{dblue}{$(-21.7)$}} & $59.9 \pm 1.0$ {\scriptsize\textcolor{dblue}{$(-22.1)$}} \\
    & & Rand-Top-K & $79.9 \pm 0.5$ {\scriptsize\textcolor{dblue}{$(-2.8)$}}  & $76.2 \pm 0.4$ {\scriptsize\textcolor{dblue}{$(-6.5)$}}  & $61.5 \pm 2.9$ {\scriptsize\textcolor{dblue}{$(-21.2)$}} & $78.8 \pm 1.2$ {\scriptsize\textcolor{dblue}{$(-3.2)$}}  & $75.4 \pm 0.6$ {\scriptsize\textcolor{dblue}{$(-6.6)$}}  & $63.3 \pm 1.2$ {\scriptsize\textcolor{dblue}{$(-18.7)$}} \\
    & & ADC        & $78.0 \pm 0.3$ {\scriptsize\textcolor{dblue}{$(-4.7)$}}  & $68.4 \pm 0.4$ {\scriptsize\textcolor{dblue}{$(-14.3)$}} & $53.7 \pm 1.0$ {\scriptsize\textcolor{dblue}{$(-29.0)$}} & $71.7 \pm 1.6$ {\scriptsize\textcolor{dblue}{$(-10.3)$}} & $68.0 \pm 1.8$ {\scriptsize\textcolor{dblue}{$(-14.0)$}} & $63.4 \pm 1.6$ {\scriptsize\textcolor{dblue}{$(-18.6)$}} \\
    \cmidrule(lr){3-9}
    & & \method{} (Ours) & $\mathbf{83.0 \pm 0.0}$ {\scriptsize\textcolor{dblue}{$\mathbf{(+0.3)}$}} & $\mathbf{82.7 \pm 0.2}$ {\scriptsize\textcolor{dblue}{$\mathbf{(+0.0)}$}} & $\mathbf{81.4 \pm 0.3}$ {\scriptsize\textcolor{dblue}{$\mathbf{(-1.3)}$}} & $\mathbf{82.7 \pm 0.3}$ {\scriptsize\textcolor{dblue}{$\mathbf{(+0.7)}$}} & $\mathbf{82.7 \pm 0.4}$ {\scriptsize\textcolor{dblue}{$\mathbf{(+0.7)}$}} & $\mathbf{81.1 \pm 0.2}$ {\scriptsize\textcolor{dblue}{$\mathbf{(-0.9)}$}} \\
    \cmidrule(lr){2-9}

    & \multirow{5}{*}{\textbf{Food101}}
    & No Compr. ($R=1$) & \multicolumn{3}{c}{$69.1 \pm 0.9$} & \multicolumn{3}{c}{$67.5 \pm 0.1$} \\
    \cmidrule(lr){3-9}
    & & C3-SL      & $37.0 \pm 2.2$ {\scriptsize\textcolor{dblue}{$(-32.1)$}} & $40.3 \pm 1.2$ {\scriptsize\textcolor{dblue}{$(-28.8)$}} & $32.1 \pm 1.2$ {\scriptsize\textcolor{dblue}{$(-37.0)$}} & $31.3 \pm 3.0$ {\scriptsize\textcolor{dblue}{$(-36.2)$}} & $31.3 \pm 3.0$ {\scriptsize\textcolor{dblue}{$(-36.2)$}} & $31.3 \pm 3.0$ {\scriptsize\textcolor{dblue}{$(-36.2)$}} \\
    & & Rand-Top-K & $66.5 \pm 1.5$ {\scriptsize\textcolor{dblue}{$(-2.6)$}}  & $60.2 \pm 0.2$ {\scriptsize\textcolor{dblue}{$(-8.9)$}}  & $35.7 \pm 4.0$ {\scriptsize\textcolor{dblue}{$(-33.4)$}} & $65.1 \pm 1.2$ {\scriptsize\textcolor{dblue}{$(-2.4)$}}  & $60.3 \pm 0.7$ {\scriptsize\textcolor{dblue}{$(-7.2)$}}  & $43.1 \pm 2.6$ {\scriptsize\textcolor{dblue}{$(-24.4)$}} \\
    & & ADC        & $63.3 \pm 1.4$ {\scriptsize\textcolor{dblue}{$(-5.8)$}}  & $54.6 \pm 1.6$ {\scriptsize\textcolor{dblue}{$(-14.5)$}} & $44.0 \pm 1.3$ {\scriptsize\textcolor{dblue}{$(-25.1)$}} & $56.2 \pm 1.3$ {\scriptsize\textcolor{dblue}{$(-11.3)$}} & $51.8 \pm 1.5$ {\scriptsize\textcolor{dblue}{$(-15.7)$}} & $47.1 \pm 1.1$ {\scriptsize\textcolor{dblue}{$(-20.4)$}} \\
    \cmidrule(lr){3-9}
    & & \method{} (Ours) & $\mathbf{70.3 \pm 1.2}$ {\scriptsize\textcolor{dblue}{$\mathbf{(+1.2)}$}} & $\mathbf{69.6 \pm 1.2}$ {\scriptsize\textcolor{dblue}{$\mathbf{(+0.5)}$}} & $\mathbf{65.3 \pm 1.7}$ {\scriptsize\textcolor{dblue}{$\mathbf{(-3.8)}$}} & $\mathbf{68.7 \pm 0.5}$ {\scriptsize\textcolor{dblue}{$\mathbf{(+1.2)}$}} & $\mathbf{68.1 \pm 0.9}$ {\scriptsize\textcolor{dblue}{$\mathbf{(+0.6)}$}} & $\mathbf{64.3 \pm 1.3}$ {\scriptsize\textcolor{dblue}{$\mathbf{(-3.2)}$}} \\
    \cmidrule(lr){2-9}

    & \multirow{5}{*}{\textbf{SUN397}}
    & No Compr. ($R=1$) & \multicolumn{3}{c}{$67.0 \pm 0.6$} & \multicolumn{3}{c}{$64.5 \pm 1.2$} \\
    \cmidrule(lr){3-9}
    & & C3-SL      & $48.0 \pm 1.1$ {\scriptsize\textcolor{dblue}{$(-19.0)$}} & $41.3 \pm 2.1$ {\scriptsize\textcolor{dblue}{$(-25.7)$}} & $31.6 \pm 2.1$ {\scriptsize\textcolor{dblue}{$(-35.4)$}} & $41.5 \pm 1.2$ {\scriptsize\textcolor{dblue}{$(-23.0)$}} & $41.5 \pm 1.2$ {\scriptsize\textcolor{dblue}{$(-23.0)$}} & $41.5 \pm 1.2$ {\scriptsize\textcolor{dblue}{$(-23.0)$}} \\
    & & Rand-Top-K & $66.3 \pm 0.7$ {\scriptsize\textcolor{dblue}{$(-0.7)$}}  & $62.9 \pm 1.0$ {\scriptsize\textcolor{dblue}{$(-4.1)$}}  & $48.5 \pm 3.1$ {\scriptsize\textcolor{dblue}{$(-18.5)$}} & $64.6 \pm 0.7$ {\scriptsize\textcolor{dblue}{$(+0.1)$}} & $61.2 \pm 1.4$ {\scriptsize\textcolor{dblue}{$(-3.3)$}} & $50.0 \pm 0.4$ {\scriptsize\textcolor{dblue}{$(-14.5)$}} \\
    & & ADC        & $62.2 \pm 0.5$ {\scriptsize\textcolor{dblue}{$(-4.8)$}}  & $52.9 \pm 1.1$ {\scriptsize\textcolor{dblue}{$(-14.1)$}} & $39.0 \pm 1.7$ {\scriptsize\textcolor{dblue}{$(-28.0)$}} & $55.5 \pm 0.6$ {\scriptsize\textcolor{dblue}{$(-9.0)$}}  & $52.1 \pm 0.8$ {\scriptsize\textcolor{dblue}{$(-12.4)$}} & $47.5 \pm 1.4$ {\scriptsize\textcolor{dblue}{$(-17.0)$}} \\
    \cmidrule(lr){3-9}
    & & \method{} (Ours) & $\mathbf{67.4 \pm 0.4}$ {\scriptsize\textcolor{dblue}{$\mathbf{(+0.4)}$}} & $\mathbf{66.7 \pm 0.3}$ {\scriptsize\textcolor{dblue}{$\mathbf{(-0.3)}$}} & $\mathbf{64.8 \pm 0.2}$ {\scriptsize\textcolor{dblue}{$\mathbf{(-2.2)}$}} & $\mathbf{65.2 \pm 1.0}$ {\scriptsize\textcolor{dblue}{$\mathbf{(+0.7)}$}} & $\mathbf{65.0 \pm 0.5}$ {\scriptsize\textcolor{dblue}{$\mathbf{(+0.5)}$}} & $\mathbf{63.8 \pm 0.5}$ {\scriptsize\textcolor{dblue}{$\mathbf{(-0.7)}$}} \\
    \specialrule{0.8pt}{3pt}{3pt}
    \multirow{6}{*}{\rotatebox[origin=c]{90}{\textit{Local Eval.}}}
    & \multirow{5}{*}{\textbf{FEMNIST}}
    & No Compr. ($R=1$) & \multicolumn{3}{c}{$87.9 \pm 0.3$} & \multicolumn{3}{c}{$85.8 \pm 0.6$} \\
    \cmidrule(lr){3-9}
    & & C3-SL      & $85.2 \pm 0.8$ {\scriptsize\textcolor{dblue}{$(-2.7)$}} & $83.5 \pm 0.1$ {\scriptsize\textcolor{dblue}{$(-4.4)$}} & $81.4 \pm 0.5$ {\scriptsize\textcolor{dblue}{$(-6.5)$}} & $80.7 \pm 1.1$ {\scriptsize\textcolor{dblue}{$(-5.1)$}} & $39.8 \pm 4.0$ {\scriptsize\textcolor{dblue}{$(-46.0)$}} & $6.9 \pm 1.8$ {\scriptsize\textcolor{dblue}{$(-78.9)$}} \\
    & & Rand-Top-K & $87.6 \pm 0.5$ {\scriptsize\textcolor{dblue}{$(-0.3)$}}  & $87.2 \pm 0.4$ {\scriptsize\textcolor{dblue}{$(-0.7)$}}  & $86.4 \pm 0.2$ {\scriptsize\textcolor{dblue}{$(-1.5)$}} & $84.6 \pm 0.7$ {\scriptsize\textcolor{dblue}{$(-1.2)$}}  & $83.8 \pm 0.4$ {\scriptsize\textcolor{dblue}{$(-2.0)$}}  & $82.2 \pm 0.3$ {\scriptsize\textcolor{dblue}{$(-3.6)$}} \\
    & & ADC        & $86.6 \pm 1.2$ {\scriptsize\textcolor{dblue}{$(-1.3)$}}  & $81.1 \pm 0.9$ {\scriptsize\textcolor{dblue}{$(-6.8)$}} & $71.8 \pm 4.3$ {\scriptsize\textcolor{dblue}{$(-16.1)$}} & $82.4 \pm 1.4$ {\scriptsize\textcolor{dblue}{$(-3.4)$}} & $78.1 \pm 3.4$ {\scriptsize\textcolor{dblue}{$(-7.7)$}} & $73.8 \pm 5.4$ {\scriptsize\textcolor{dblue}{$(-12.0)$}} \\
    \cmidrule(lr){3-9}
    & & \method{} (Ours) & $\mathbf{88.1 \pm 0.3}$ {\scriptsize\textcolor{dblue}{$\mathbf{(+0.2)}$}} & $\mathbf{88.0 \pm 0.1}$ {\scriptsize\textcolor{dblue}{$\mathbf{(+0.1)}$}} & $\mathbf{87.9 \pm 0.5}$ {\scriptsize\textcolor{dblue}{$\mathbf{(+0.0)}$}} & $\mathbf{85.9 \pm 0.4}$ {\scriptsize\textcolor{dblue}{$\mathbf{(+0.1)}$}} & $\mathbf{85.5 \pm 0.9}$ {\scriptsize\textcolor{dblue}{$\mathbf{(-0.3)}$}} & $\mathbf{84.6 \pm 1.0}$ {\scriptsize\textcolor{dblue}{$\mathbf{(-1.2)}$}} \\
    \bottomrule
  \end{tabular}%
  }
\end{table}

\begin{figure}[htbp]
    \centering
    \begin{subfigure}[b]{0.49\textwidth}
        \centering
        \includegraphics[width=\textwidth]{figures/final/comm_cifar100_5.pdf}
        \caption{\textbf{CIFAR-100} / $N=5$}
        \label{fig:cifar100_5}
    \end{subfigure}
    \hfill
    \begin{subfigure}[b]{0.49\textwidth}
        \centering
        \includegraphics[width=\textwidth]{figures/final/comm_cifar100_25.pdf}
        \caption{\textbf{CIFAR-100} / $N=25$}
        \label{fig:cifar100_25}
    \end{subfigure}

    \vspace{0.5cm} 

    \begin{subfigure}[b]{0.49\textwidth}
        \centering
        \includegraphics[width=\textwidth]{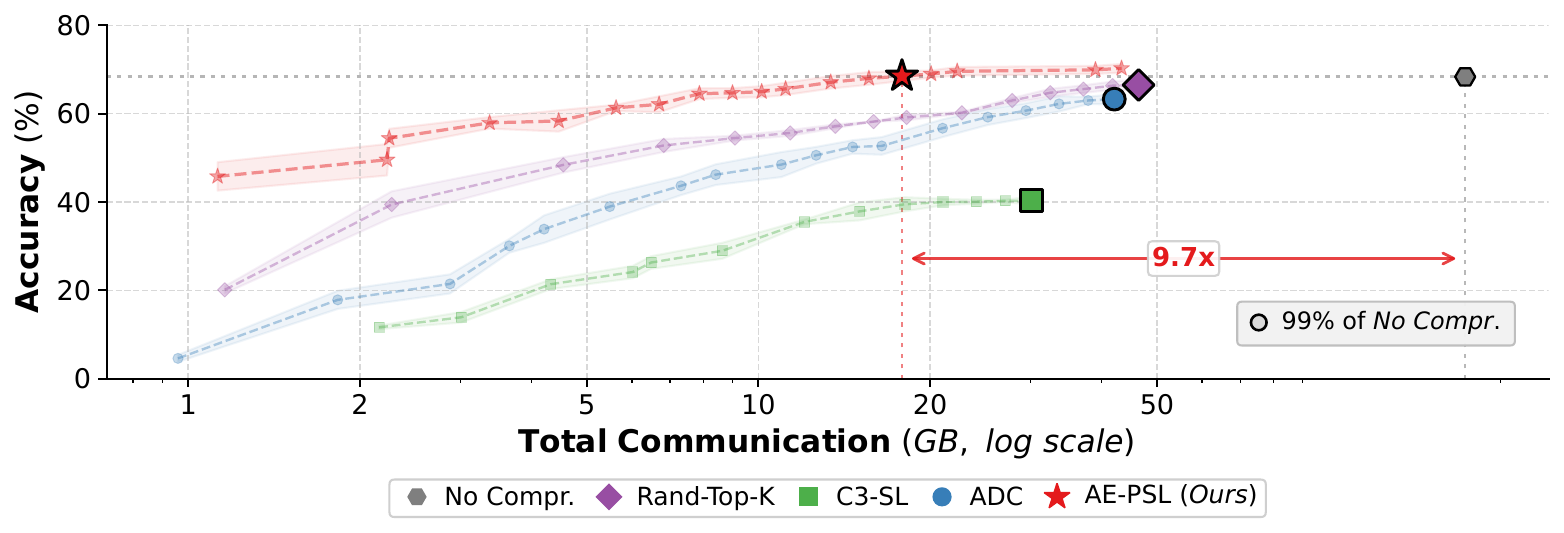}
        \caption{\textbf{Food101} / $N=5$ }
        \label{fig:food101_5}
    \end{subfigure}
    \hfill
    \begin{subfigure}[b]{0.49\textwidth}
        \centering
        \includegraphics[width=\textwidth]{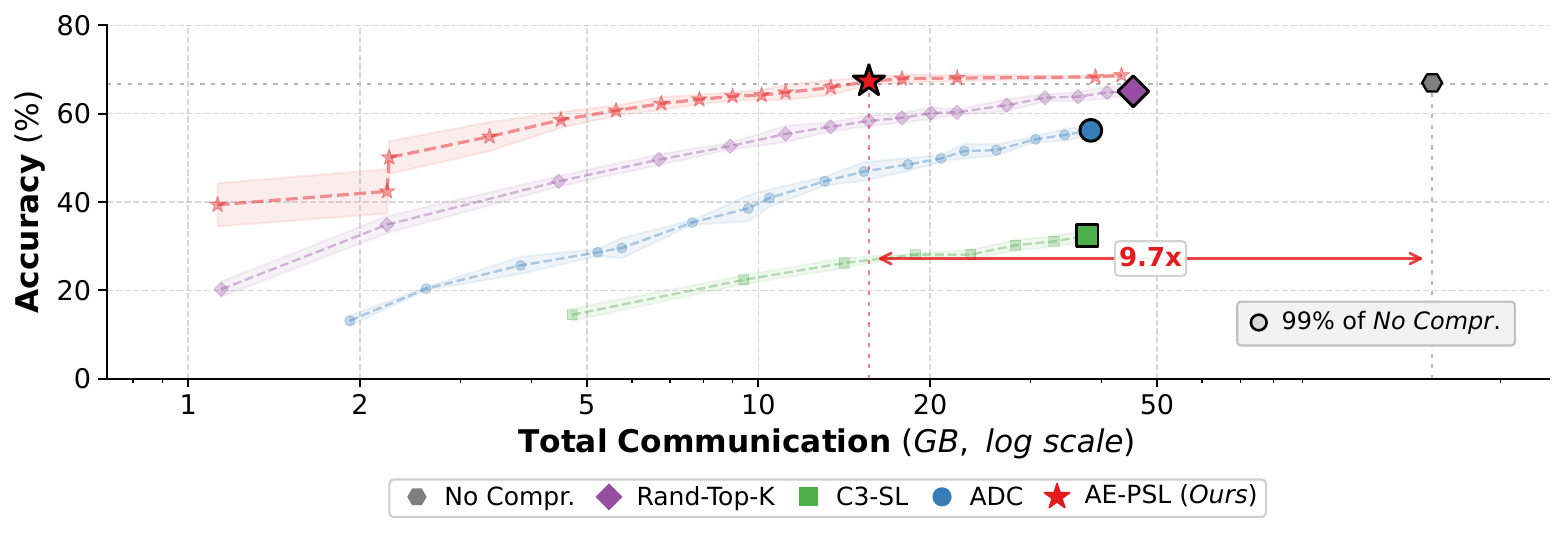}
        \caption{\textbf{Food101} / $N=25$}
        \label{fig:food101_25}
    \end{subfigure}
    
    \vspace{0.5cm}
    
        \begin{subfigure}[b]{0.49\textwidth}
        \centering
        \includegraphics[width=\textwidth]{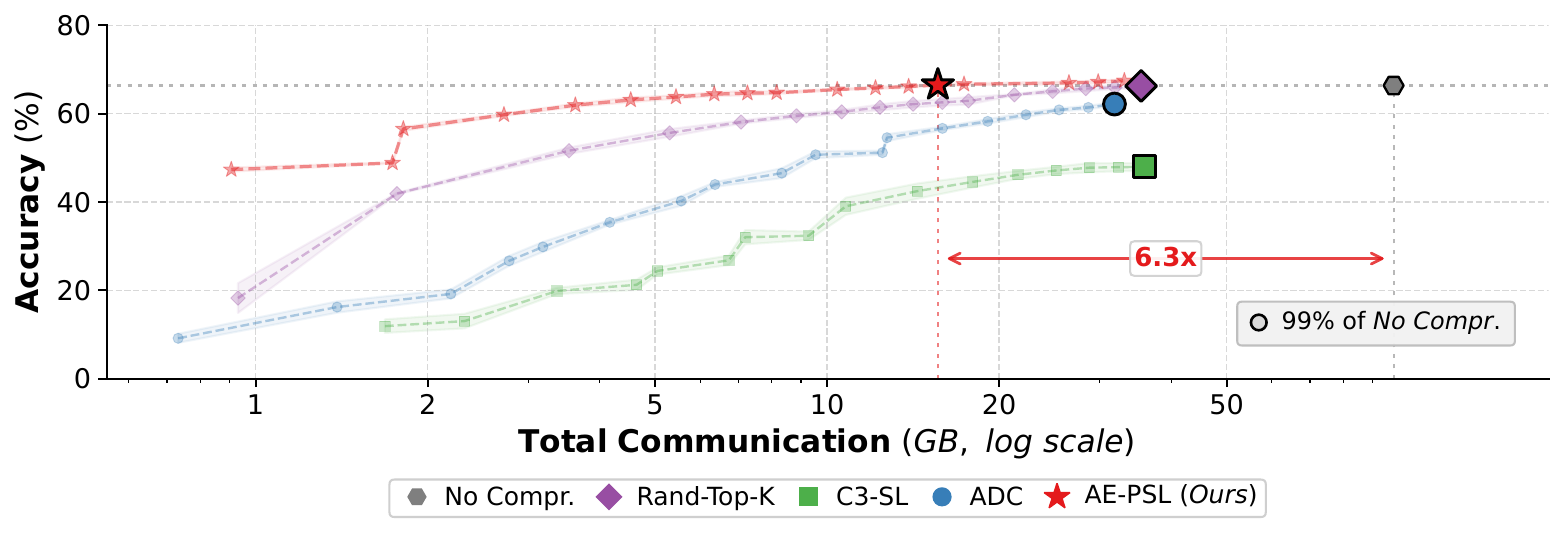}
        \caption{\textbf{SUN397} / $N=5$}
        \label{fig:sun397_5}
    \end{subfigure}
    \hfill
    \begin{subfigure}[b]{0.49\textwidth}
        \centering
        \includegraphics[width=\textwidth]{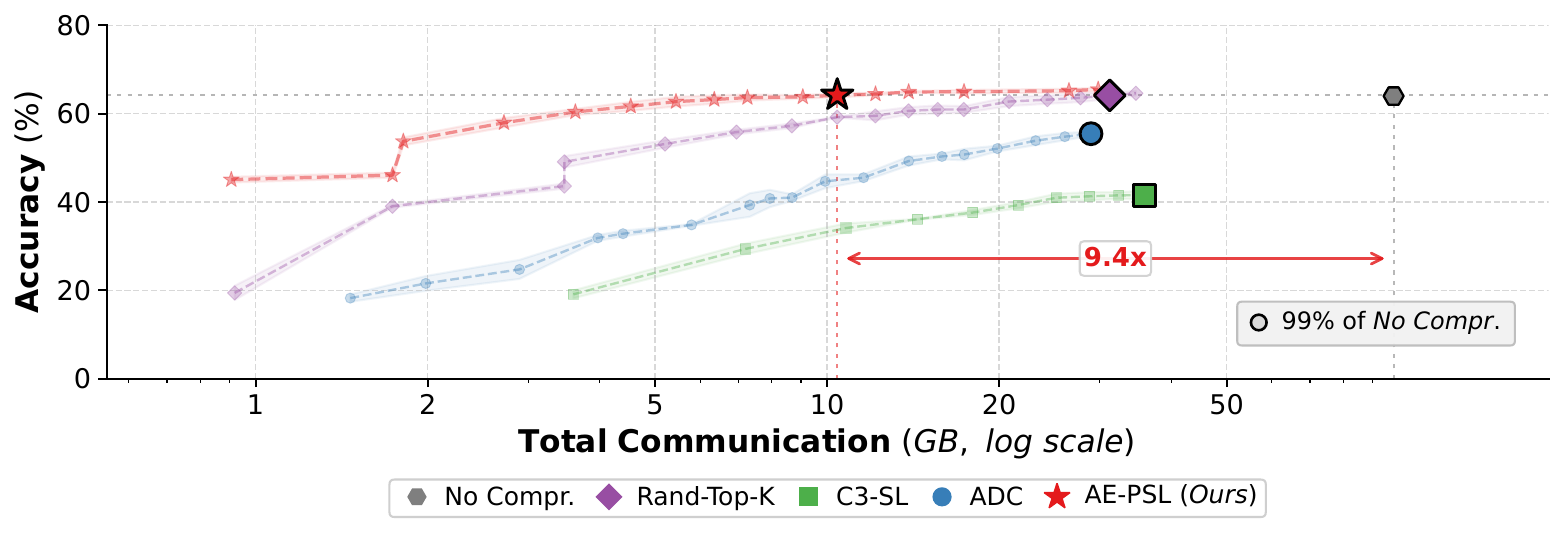}
        \caption{\textbf{SUN397} / $N=25$}
        \label{fig:sun397_25}
    \end{subfigure}
    
    \vspace{0.5cm}
    
    \begin{subfigure}[b]{0.49\textwidth}
        \centering
        \includegraphics[width=\textwidth]{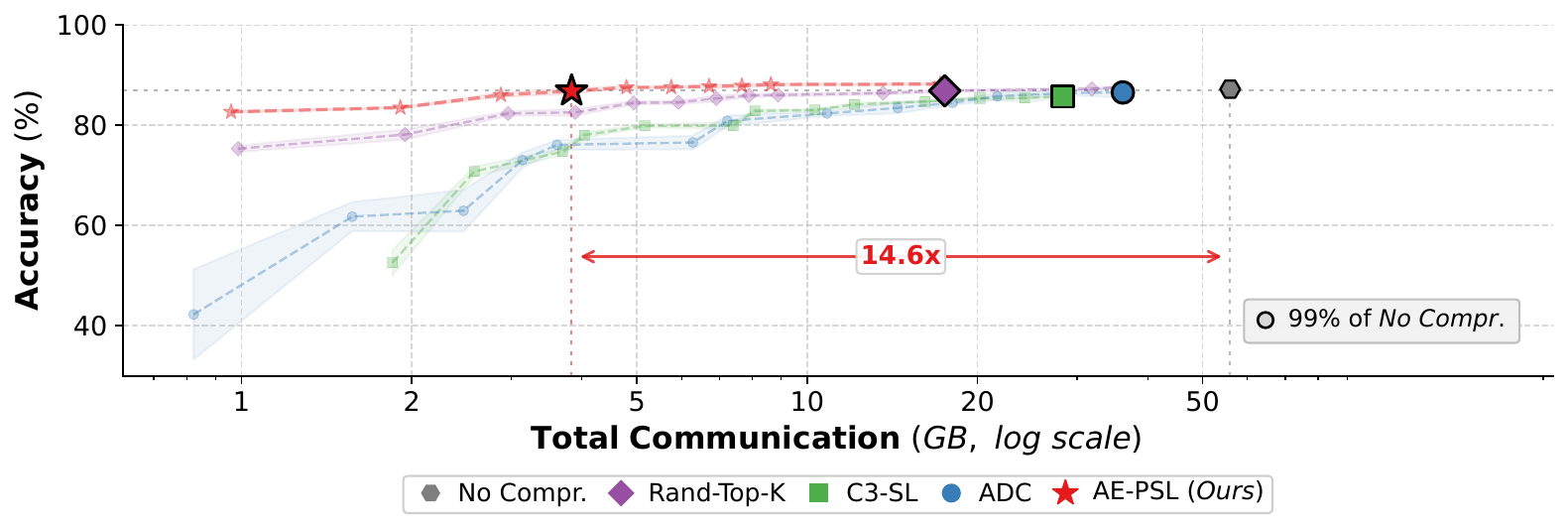}
        \caption{\textbf{FEMNIST} / $N=5$}
        \label{fig:femnist_5}
    \end{subfigure}
    \hfill
    \begin{subfigure}[b]{0.49\textwidth}
        \centering
        \includegraphics[width=\textwidth]{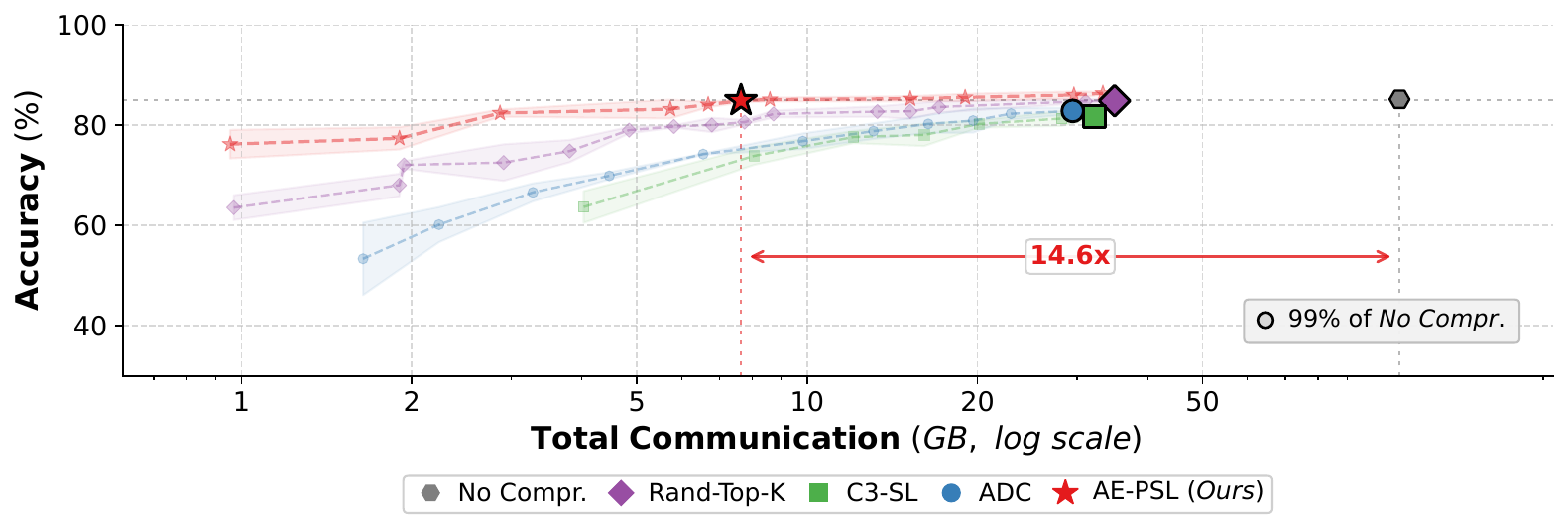}
        \caption{\textbf{FEMNIST} / $N=25$}
        \label{fig:femnist_25}
    \end{subfigure}
    \caption{Accuracy--communication trade-off using ViT-B/32 for $N=5$ and $N=25$ across datasets. We report average accuracy across $3$ seeds for different communication volumes (GB); markers indicate when a method reaches >$95\%$ of peak performance.} \label{fig:total_communication_comparison}
\end{figure}

\vspace{3pt}\noindent\textit{\textbf{Performance under Fixed Communication Budget.}}
Here, we further evaluate the total communication overhead across clients on all datasets, shown in Fig. \ref{fig:total_communication_comparison}. The observed patterns remain consistent across all datasets.

\vspace{3pt}\noindent\textit{\textbf{Client-side Computational Overhead.}}
Here we evaluate the total communication overhead across clients. In Fig. \ref{fig:accuracy_vs_gflops_all} plots for all datasets and client counts are presented.

\begin{figure}[htbp]
    \centering
    
    \begin{subfigure}[b]{0.95\textwidth}
        \centering
        \includegraphics[width=\textwidth]{figures/final/gflops_cifar100_5.pdf}
        \caption{\textbf{CIFAR-100} / $N=5$}
        \label{fig:cifar100_5}
    \end{subfigure}
    \vspace{0.5cm}

    \begin{subfigure}[b]{0.95\textwidth}
        \centering
        \includegraphics[width=\textwidth]{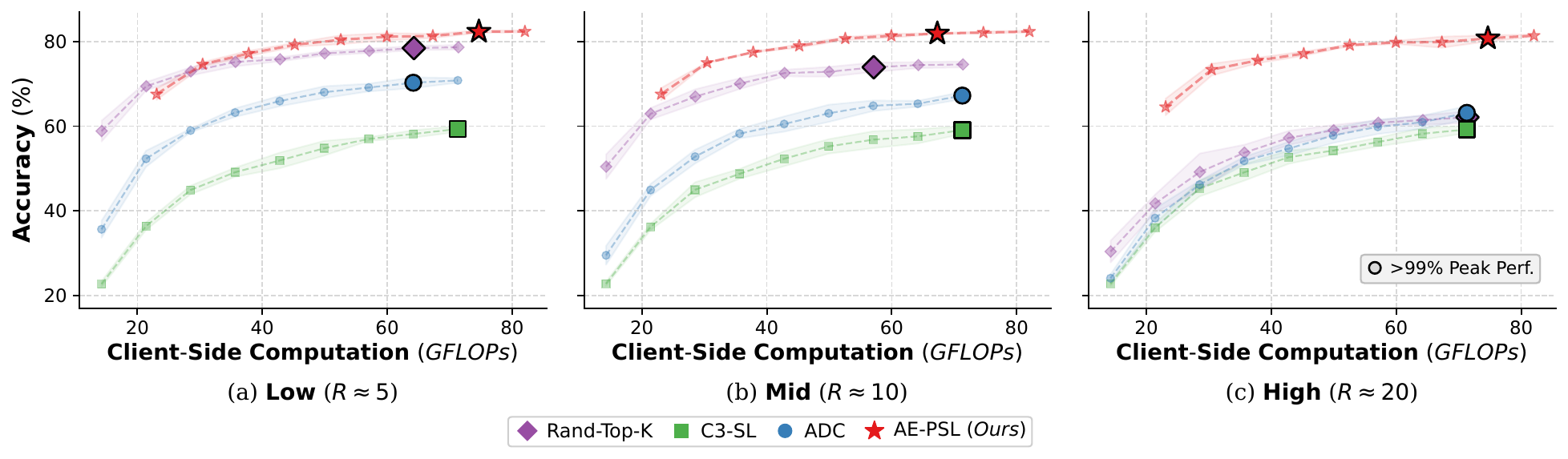}
        \caption{\textbf{CIFAR-100} / $N=25$}
        \label{fig:cifar100_25}
    \end{subfigure}
    \vspace{0.5cm}

    \begin{subfigure}[b]{0.95\textwidth}
        \centering
        \includegraphics[width=\textwidth]{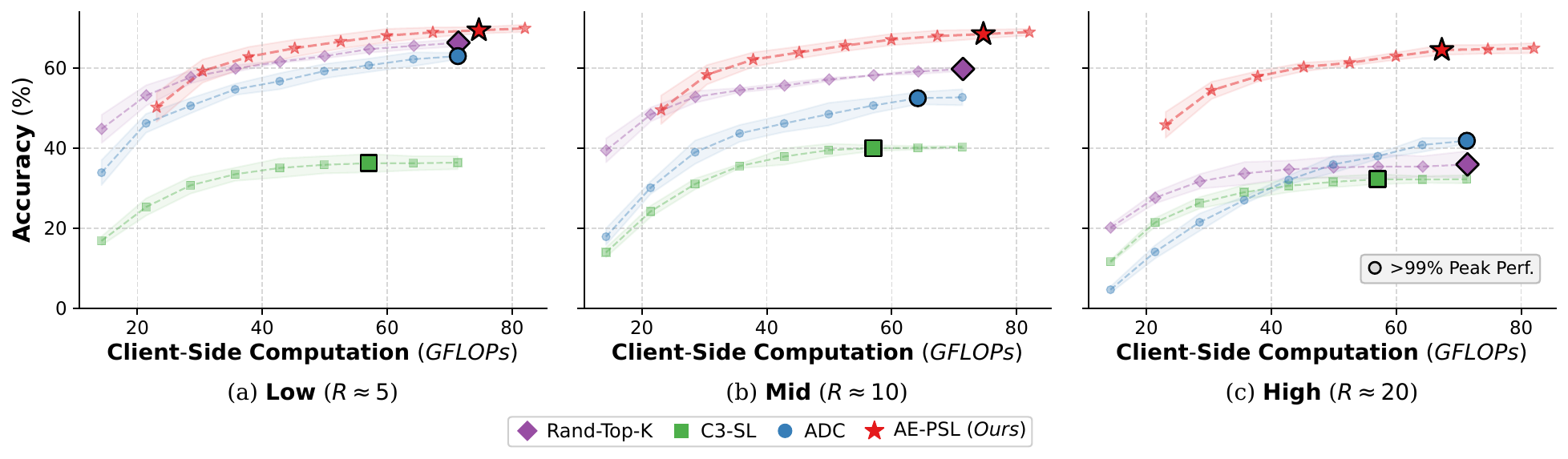}
        \caption{\textbf{Food101} / $N=5$ }
        \label{fig:food101_5}
    \end{subfigure}
    \vspace{0.5cm}

    \begin{subfigure}[b]{0.95\textwidth}
        \centering
        \includegraphics[width=\textwidth]{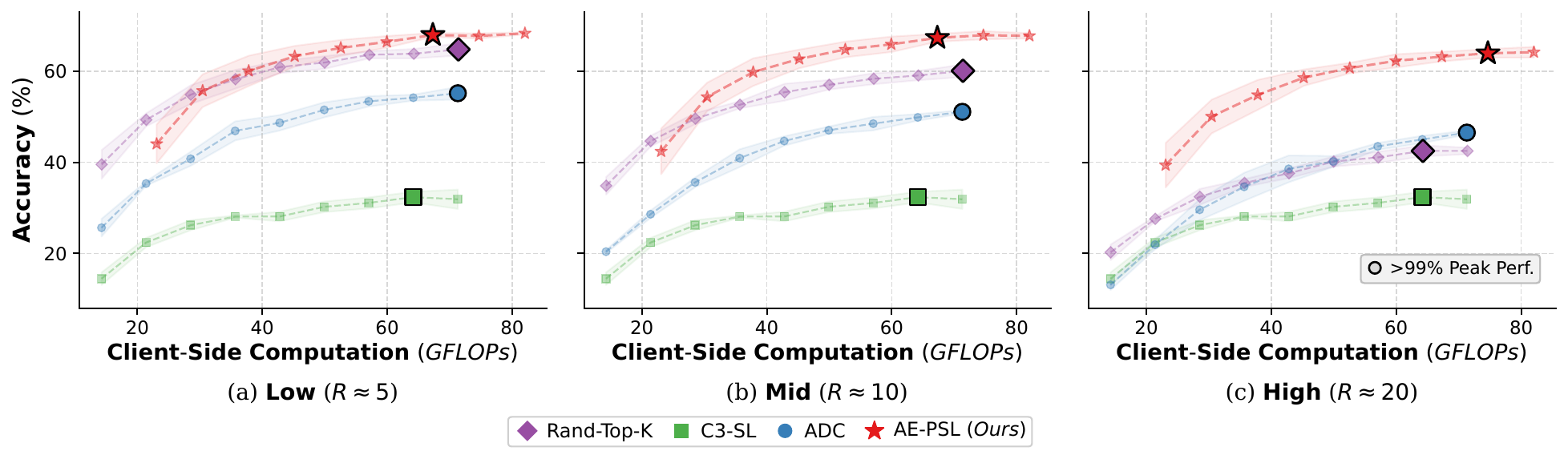}
        \caption{\textbf{Food101} / $N=25$}
        \label{fig:food101_25}
    \end{subfigure}

    \caption{Accuracy--client-side compute trade-off using ViT-B/32 for $N\in\{5,25\}$ and $R\in\{5,10,20\}$ across datasets. We report average accuracy across $3$ seeds for cumulative client-side computational overhead (GFLOPs) across $R$ levels; markers indicate when a method reaches >$95\%$ of peak performance. (Continued on next page)}
    \label{fig:accuracy_vs_gflops_all}
\end{figure}

\begin{figure}[htbp]
    \centering
    \ContinuedFloat 
    
    \begin{subfigure}[b]{0.95\textwidth}
        \centering
        \includegraphics[width=\textwidth]{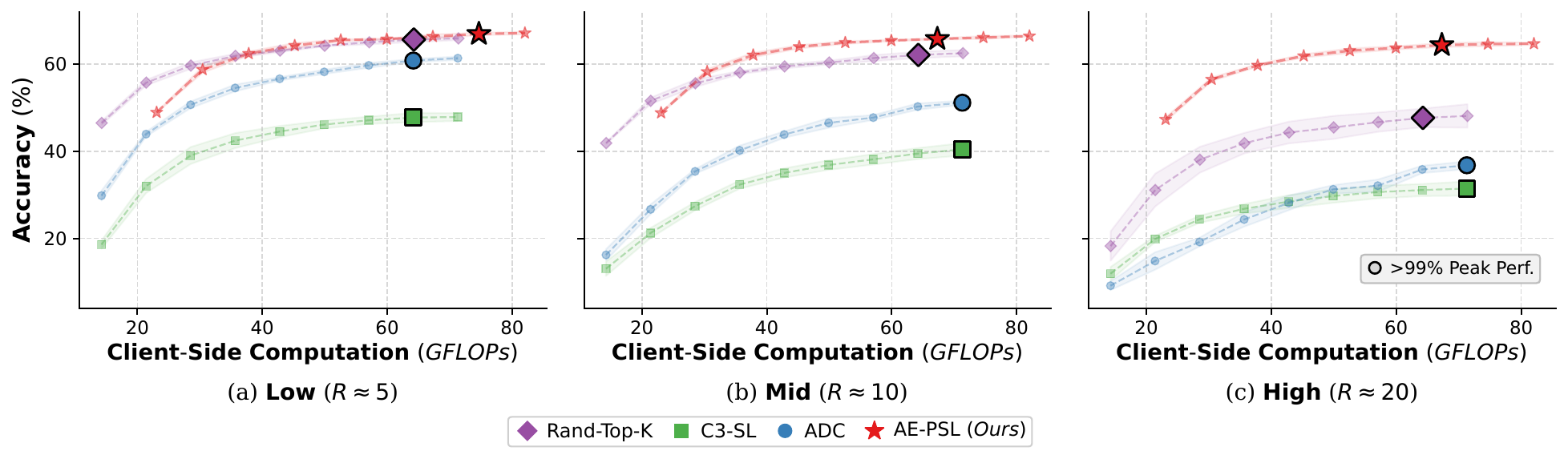}
        \caption{\textbf{SUN397} / $N=5$}
        \label{fig:sun397_5}
    \end{subfigure}
    \vspace{0.5cm}

    \begin{subfigure}[b]{0.95\textwidth}
        \centering
        \includegraphics[width=\textwidth]{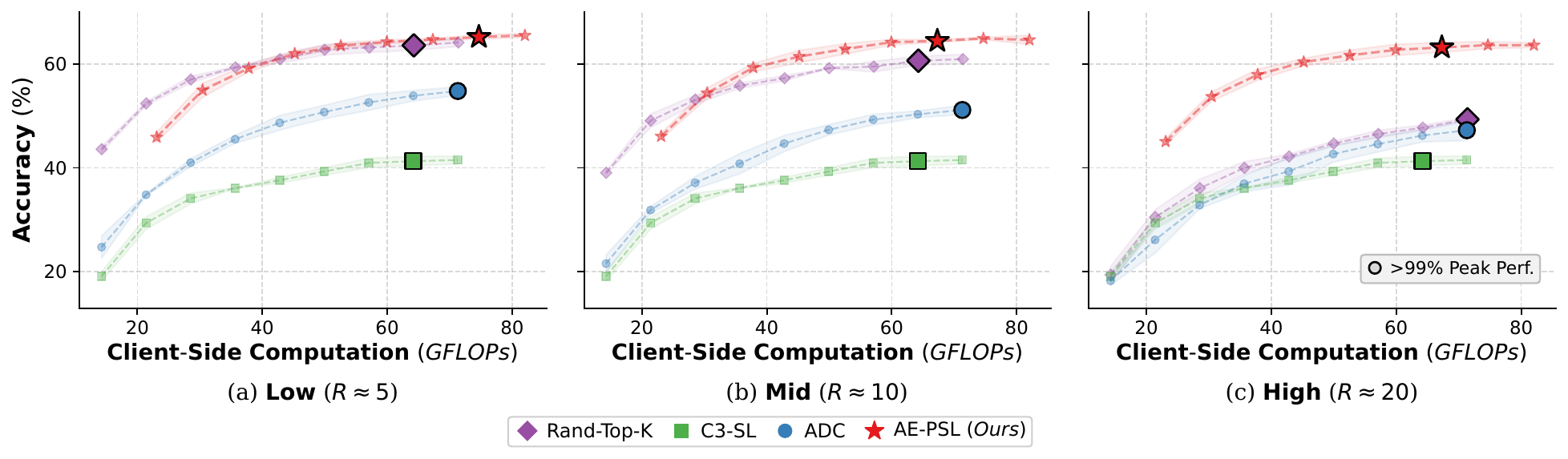}
        \caption{\textbf{SUN397} / $N=25$}
        \label{fig:sun397_25}
    \end{subfigure}
    \vspace{0.5cm}
    
    \begin{subfigure}[b]{0.95\textwidth}
        \centering
        \includegraphics[width=\textwidth]{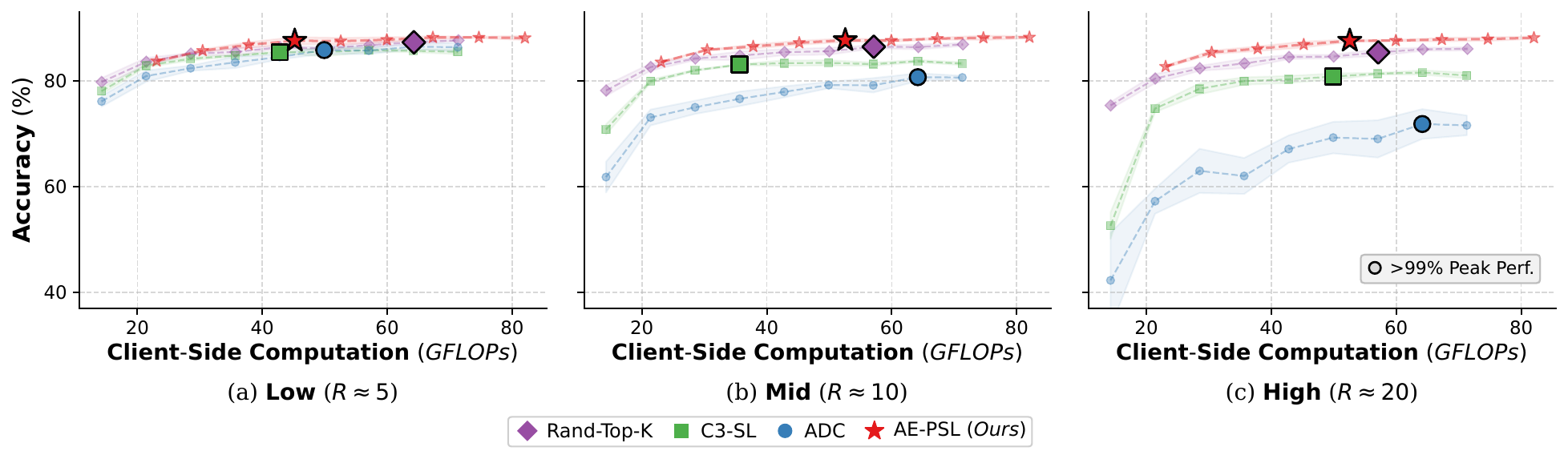}
        \caption{\textbf{FEMNIST} / $N=5$}
        \label{fig:femnist_5}
    \end{subfigure}
    \vspace{0.5cm}

    \begin{subfigure}[b]{0.95\textwidth}
        \centering
        \includegraphics[width=\textwidth]{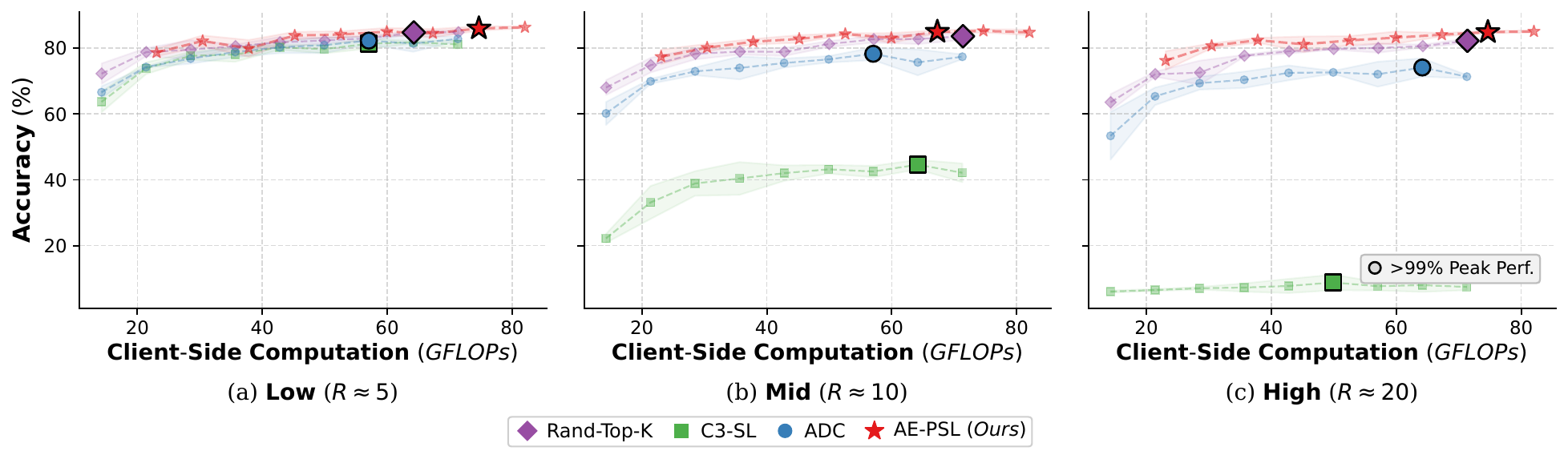}
        \caption{\textbf{FEMNIST} / $N=25$}
        \label{fig:femnist_25}
    \end{subfigure}

    \caption{Accuracy--client-side compute trade-off using ViT-B/32 for $N\in\{5,25\}$ and $R\in\{5,10,20\}$ across datasets. We report average accuracy across $3$ seeds for cumulative client-side computational overhead (GFLOPs) across $R$ levels; markers indicate when a method reaches >$95\%$ of peak performance.}
\end{figure}

\newpage

\subsection{Additional Analysis of Local Evaluation} \label{app:local_eval}
As shown in Table~\ref{tab:femnist_ae_eval}, downstream performance is largely preserved when a randomly initialized AE is introduced, provided the module remains active during inference. Conversely, bypassing the AE at test time precipitates a severe degradation in accuracy. This occurs because the primary model and the AE co-adapt during DFT; removing the compression module subsequently induces a substantial feature distribution misalignment. 

\begin{table}[!h]
  \centering \scriptsize
  \setlength{\tabcolsep}{3.8pt}
  \renewcommand{\arraystretch}{1.13}
  \caption{Effect of the AE module during FEMNIST \textit{local evaluation}. We report average accuracy across $3$ seeds at $R\approx15$, with the AE active at inference time (\textit{w/ AE}) and removed at inference time (\textit{w/o AE}); deltas denote relative performance with respect to the \textit{no compression} setting ($R=1$).}
  \label{tab:femnist_ae_eval}
  \resizebox{\linewidth}{!}{%
  \begin{tabular}{l cc cc c}
    \toprule
    \multirow{2}{*}{\textbf{Method}} &
    \multicolumn{2}{c}{$\mathbf{N=5}$} &
    \multicolumn{2}{c}{$\mathbf{N=25}$} &
    \multirow{2}{*}{\textbf{R}} \\
    \cmidrule(lr){2-3} \cmidrule(lr){4-5}
    & \textit{Local Eval. (w/ AE)} & \textit{Local Eval. (w/o AE)}
    & \textit{Local Eval. (w/ AE)} & \textit{Local Eval. (w/o AE)} & \\
    \midrule
    No Compression
    & \multicolumn{2}{c}{$87.6 \pm 0.7$}
    & \multicolumn{2}{c}{$85.4 \pm 1.3$}
    & --- \\
    \cmidrule(lr){1-6}
    AE
    & \abres{86.9}{2.0}{-0.7}
    & \abres{1.4}{1.3}{-86.2}
    & \abres{84.3}{1.8}{-1.1}
    & \abres{1.8}{3.3}{-83.6}
    & $15\times$ \\
    AE \!+\! GA
    & \abres{87.1}{1.1}{-0.5}
    & \abres{54.3}{12.3}{-33.3}
    & \abresbest{85.5}{1.1}{+0.1}
    & \abres{54.5}{11.8}{-30.9}
    & $15\times$ \\
    AE \!+\! GA \!+\! FZ
    & \abresbest{88.0}{0.6}{+0.4}
    & \abres{81.9}{0.7}{-5.7}
    & \abres{84.8}{0.9}{-0.6}
    & \abresbest{81.3}{2.8}{-4.1}
    & $15\times$ \\
    AE \!+\! GA \!+\! CSA \!+\! FZ (\method)
    & \abres{88.0}{0.7}{+0.4}
    & \abresbest{82.9}{2.2}{-4.7}
    & \abres{84.1}{0.8}{-1.3}
    & \abres{80.5}{1.6}{-4.9}
    & $15\times$ \\
    \bottomrule
  \end{tabular}%
  }
\end{table}

\newpage

\subsection{Additional Analysis of \method\ Components} \label{app:training_integration}

To gain a deeper understanding of why \method{} components improve downstream task accuracy, we measure AE reconstruction fidelity throughout DFT. We define this fine-tuning reconstruction error as $\mathcal{E}_{\text{rec}} = \|\mathbf{H}_n - \widehat{\mathbf{H}}_n\|_2^2$, which computes the mean squared error between the original client-side patch activations $\mathbf{H}_n$ and the server-side reconstructions $\widehat{\mathbf{H}}_n$, averaged across clients for a given batch. We distinguish $\mathcal{E}_{\text{rec}}$ from the optimization objective used during the GA phase, as $\mathcal{E}_{\text{rec}}$ serves strictly as an evaluation metric during DFT. 

\begin{figure}[!h]
    \centering \small\includegraphics[width=0.8\linewidth]{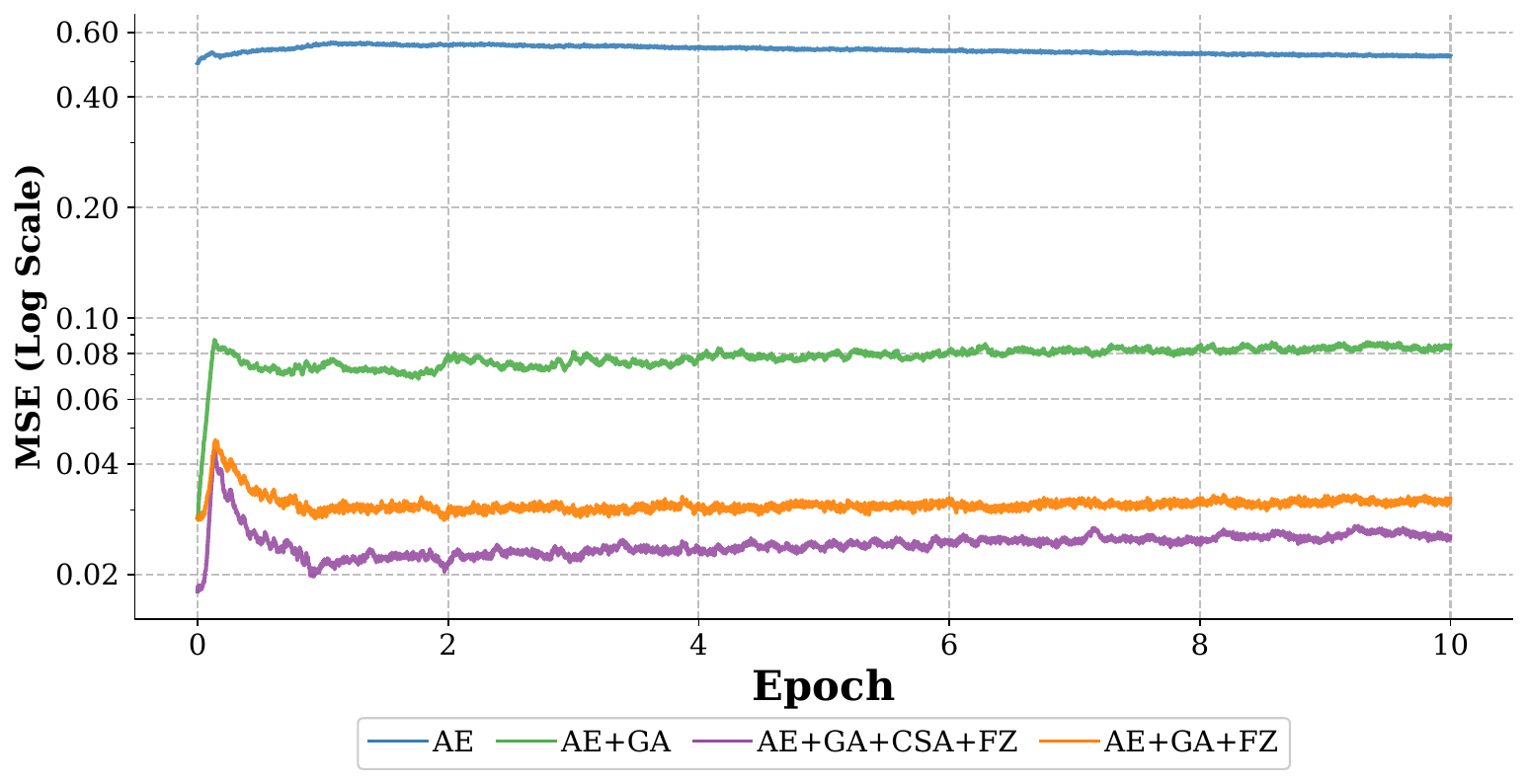}
    \caption{Per-client averaged reconstruction error ($\mathcal{E}_{\text{rec}}$) during DFT for $N=5$ on CIFAR-100.}
    \label{fig:mse_warmup}
\end{figure}

As evidenced in Fig. \ref{fig:mse_warmup}, inserting a randomly initialized AE causes high initial $\mathcal{E}_{\text{rec}}$. The addition of the GA phase (Sec.~\ref{sec:GA}) significantly reduces the initial error $\mathcal{E}_{\text{rec}}$, however rapidly increases as DFT progresses if the AE parameters are left unfrozen. By freezing (FZ) the AE during DFT, $\mathcal{E}_{\text{rec}}$ is relatively stable. This indicates that updating the AE weights concurrently with the task optimization (via Cross Entropy loss) yields gradients that counter-intuitively degrade reconstruction fidelity. We hypothesize that while the primary model weights and the AE may co-adapt—causing the latent space to drift—this adaptation is detrimental under the \textit{global} evaluation, where the AE is explicitly bypassed during inference. Finally, the application of CSA (Sec.~\ref{sec:CSA}) establishes an even lower $\mathcal{E}_{\text{rec}}$ at the onset of DFT, directly correlating with the increased downstream task accuracy observed in our main results.

\newpage
\subsection{Alternative AE Alignment Strategies.} \label{app:alternative_alignment}
Beyond the two-stage alignment procedure detailed in Sec.~\ref{sec:two-stage-AEA}, we investigated several alternative training strategies aimed directly at minimizing $\mathcal{E}_{\text{rec}}$ during fine-tuning.

\begin{figure}[!h]
    \centering \small
    \includegraphics[width=0.8\linewidth]{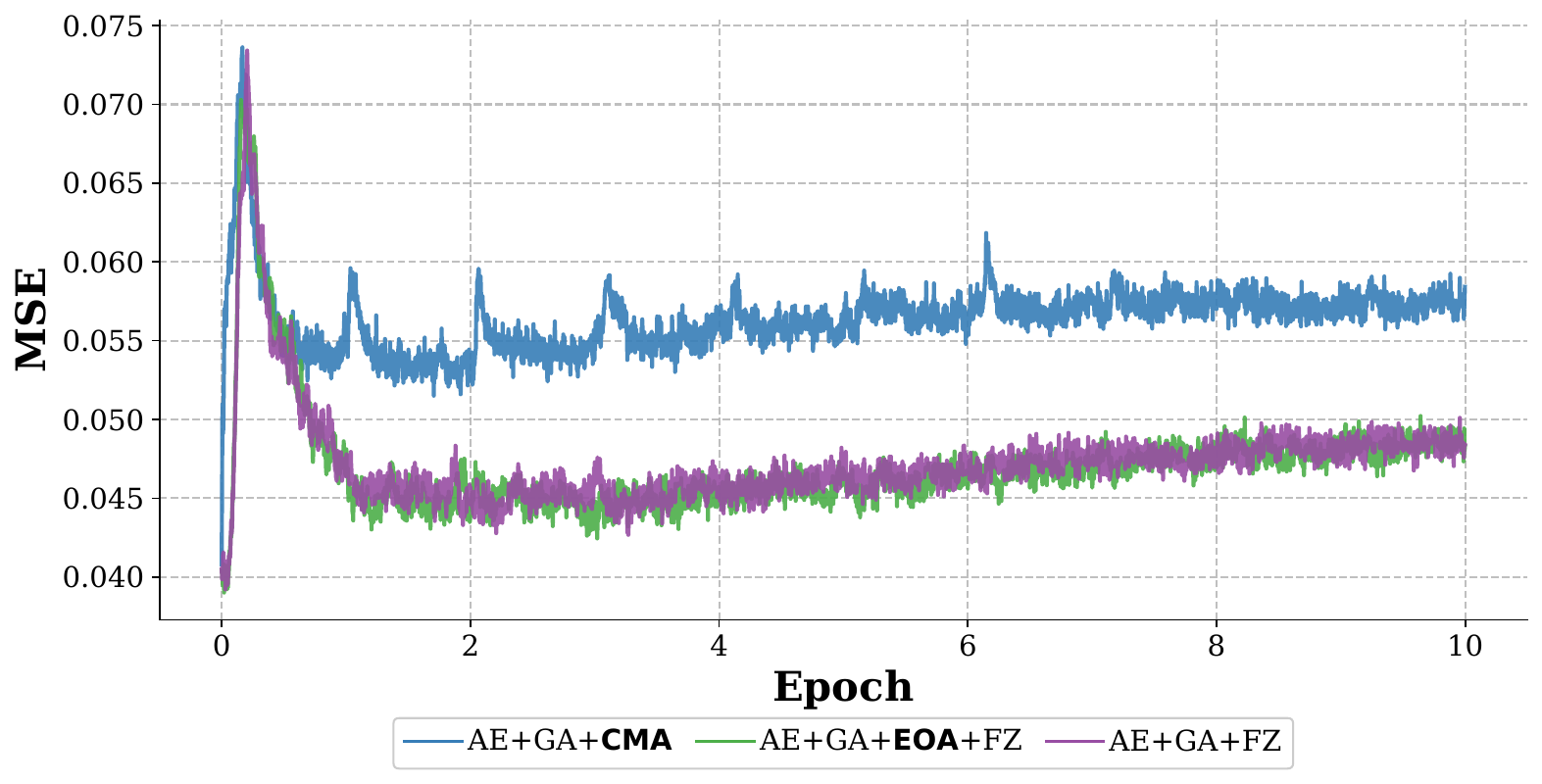}
    \caption{Per-client averaged reconstruction error ($\mathcal{E}_{\text{rec}}$) during DFT for $N=5$ on CIFAR-100.}
    \label{fig:additional_training}
\end{figure}

\vspace{3pt} \noindent \textbf{Encoder-Only Client-Specific Alignment (EO-CSA).}
To eliminate the communication overhead associated with server-side decoder aggregation of standard CSA, we evaluated an Encoder-Only Client-Specific Alignment strategy. In this configuration, only the client-side encoder $E_{\boldsymbol{\phi}_n}$ is updated during the local warm-up phase, while the server-side decoder $D_{\boldsymbol{\psi}}$ remains strictly frozen. As demonstrated in Fig. \ref{fig:additional_training} and Table \ref{tab:alternative}, EO-CSA yields negligible AE adaptation; the encoder struggles to map local data distributions to the existing latent space effectively without corresponding weight updates in the decoder.

\begin{table}[!b]
  \centering \scriptsize
  \setlength{\tabcolsep}{4.5pt}
  \renewcommand{\arraystretch}{1.12}
  \caption{Effect of alternative AE alignment strategies: Encoder-Only Client-Specific Alignment (EO-CSA), Concurrent MSE Alignment (CMA), and iterative Client-Specific Alignment ($\ell$-CSA). Experiments are conducted with ViT-B/32 under \textit{global} evaluation for $R=15$ with $N=5$ on CIFAR100. We report average accuracy across $3$ seeds.}
  \label{tab:alternative}
  \resizebox{0.6\linewidth}{!}{%
  \begin{tabular}{l c}
    \toprule
    \textbf{AE Alignment Strategy} & \textbf{Accuracy (\%)} \\
    \midrule
    AE \!+\! GA \!+\! FZ
    & $79.6 \pm 0.9$ \\

    AE \!+\! GA \!+\! EO-CSA \!+\! FZ
    & $79.6 \pm 0.7$ \\

    AE \!+\! GA \!+\! CMA
    & $78.1 \pm 1.1$ \\
    \midrule
    AE \!+\! GA \!+\! $1$-CSA \!+\! FZ (\method)
    & $82.2 \pm 0.7$ \\

    AE \!+\! GA \!+\! $2$-CSA \!+\! FZ
    & $82.4 \pm 0.9$ \\

    AE \!+\! GA \!+\! $3$-CSA \!+\! FZ
    & $\mathbf{82.4 \pm 0.8}$ \\
    \bottomrule
  \end{tabular}%
  }
\end{table}

\vspace{3pt} \noindent \textbf{Concurrent MSE Alignment (CMA).}
To address the drift in reconstruction error observed during DFT without freezing, we evaluate Concurrent MSE Alignment (CMA). Instead of freezing the compression module entirely, we maintained a frozen local copy of the global server-side decoder $D_{\boldsymbol{\psi}^{\text{global}}}$ on the client. During the forward pass, an auxiliary MSE loss was computed locally to update the encoder $E_{\boldsymbol{\phi}_n}$ concurrently with the primary task optimization. The regular cross entropy loss assumes the AE as an identity function in CMA. However, CMA underperformed compared to the strictly frozen AE approach, both in terms of $\mathcal{E}_{\text{rec}}$ and downstream task accuracy as visualized in Fig. \ref{fig:additional_training} and Table \ref{tab:alternative}. This performance degradation likely stems from conflicting optimization signals: the main cross entropy gradient implicitly assumes a static mapping across the split layer, while the auxiliary MSE gradient continuously perturbs the encoder's latent space. This finding highlights that adapting the decoder to the latent projections is critical for preserving reconstruction fidelity, validating our choice to utilize a client-side warm-up combined with server-side decoder aggregation.

\begin{figure}[!h]
    \centering \center
    \includegraphics[width=0.8\linewidth]{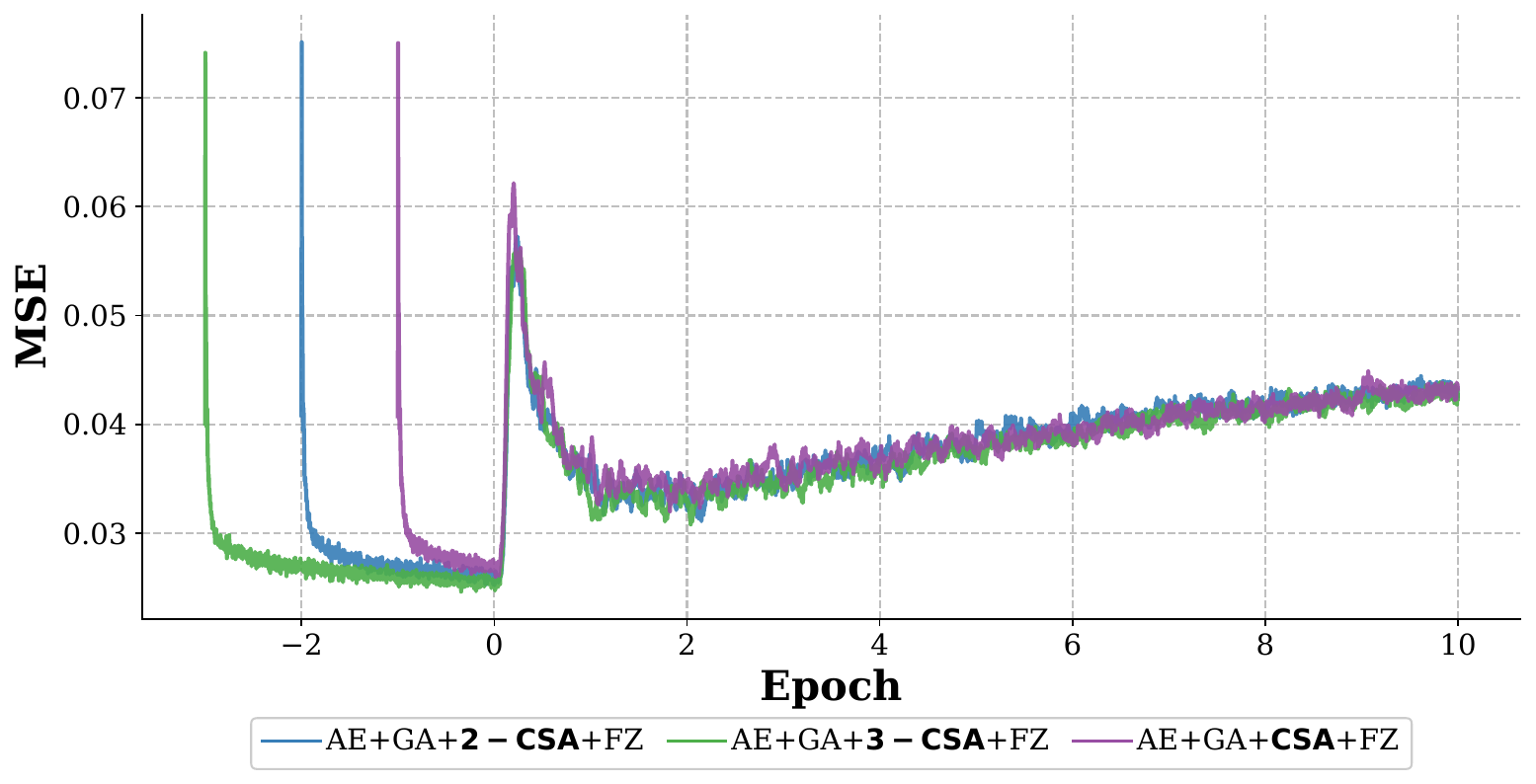}
    \caption{Per-client averaged reconstruction error ($\mathcal{E}_{\text{rec}}$) of the AE during DFT, for $N=5$ on CIFAR-100. The epochs of the $\ell$-CSA phase are visualized as negative epochs. Note that no server-side training occurs during the CSA phase.}
    \label{fig:l-CSA}
\end{figure}

\vspace{3pt} \noindent \textbf{Iterative Client-Specific Alignment ($\ell$-CSA).}
We explored extending the standard single-round CSA (equivalent to $1$-CSA) into an Iterative Client-Specific Alignment ($\ell$-CSA) process spanning multiple epochs. Under $\ell$-CSA, local client-specific alignment and subsequent server-side decoder aggregation are repeated for $\ell$ epochs, where the aggregated global decoder is redistributed to all clients at the end of each epoch. As illustrated in Fig. \ref{fig:l-CSA} and Table \ref{tab:alternative}, this iterative approach yielded a marginal reduction in reconstruction error and an insignificant increase in downstream task accuracy compared to standard $1$-CSA. Consequently, the substantial increase in client-side computational load required for $\ell$ client-side epochs renders $\ell$-CSA inefficient relative to the minimal performance gains achieved.

\newpage
\subsection{Impact of Split Layer Selection} \label{app:split_layer}
In order to analyze the optimal split layer $s$ we training multiple dedicated AEs via GA (as detailed in Sec.~\ref{sec:GA}). For each candidate layer $0 \le s \le 6$, an AE is aligned on intermediate activations extracted from the public ImageNet100 dataset. Subsequently each AE is integrated into the model at its respective split point to evaluate downstream top-1 accuracy via DFT. 

\begin{figure}[!h]
    \centering \small
    \includegraphics[width=0.9\linewidth]{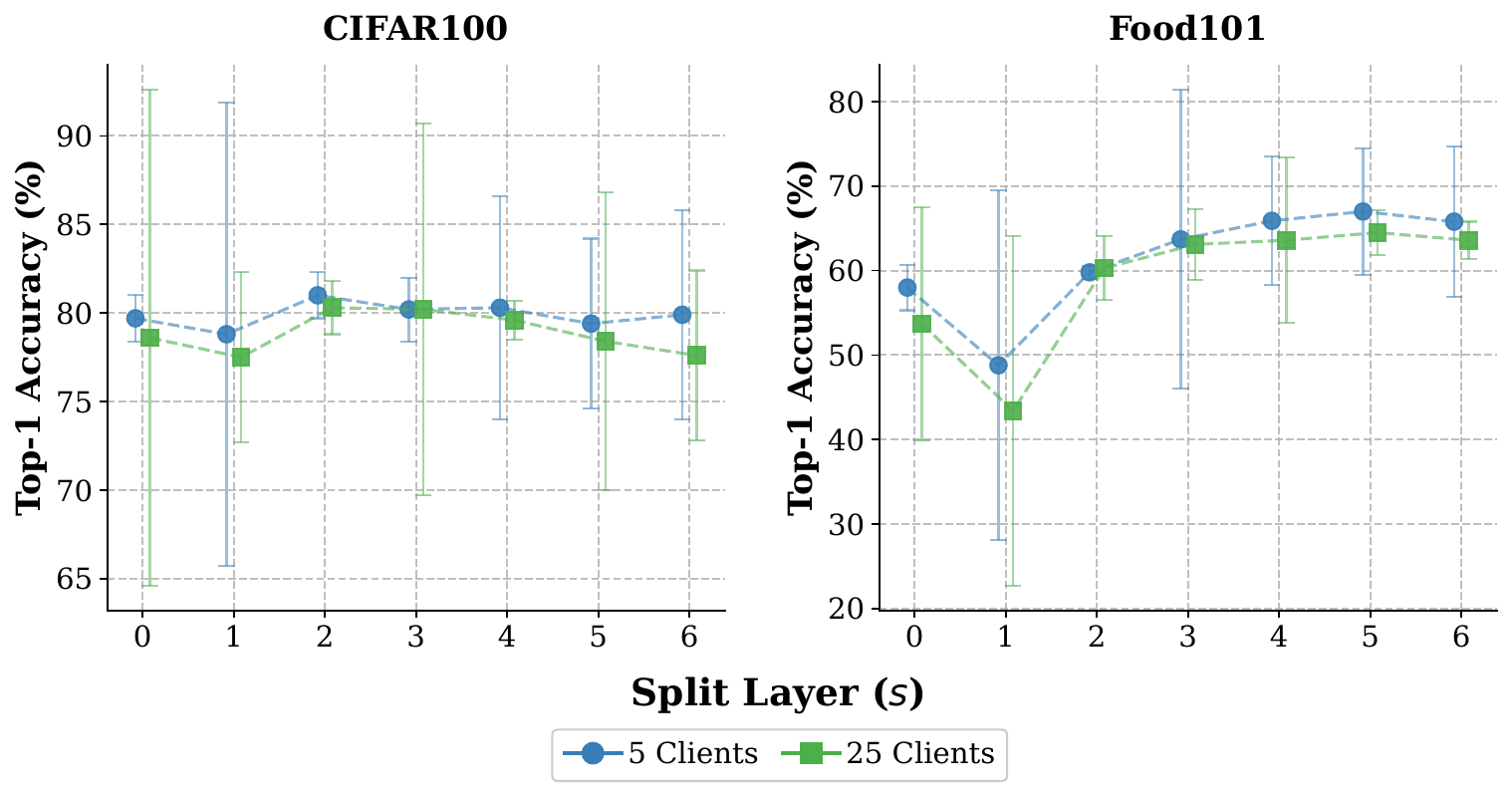}
    \caption{Accuracy of DFT with AE compression at split layer $s$, evaluated on ViT-B\textbackslash32 which consists of $12$ layers in total. Communication reduction level is fixed ($R \approx 15$). Data points are slightly offset to enhance readability of the error intervals.}
    \label{fig:split_layer}
\end{figure}

As illustrated in Fig.~\ref{fig:split_layer}, the choice of split layer significantly influences on downstream performance, showing distinct trends between datasets. Notably, for Food101, positioning the bottleneck at early layers ($s \in \{1, 2\}$) induces substantial performance degradation. On the other hand, while the average downstream accuracy for CIFAR-100 remains stable at these early layers, we observe a significant increase in variance, indicating that early split layers introduce training instability.

Configuring a lower split layer is advantageous because it offloads a larger portion of the model to the server, thereby minimizing client-side memory and compute requirements. However, our evaluation suggest that no clear universally optimal split point exists across evaluated settings. Consequently, to balance client-side computational overhead against training stability and task accuracy, $s=5$ was selected as the default configuration for all experiments.

\newpage
\subsection{AutoEncoder Architecture \& Computational Overhead Analysis} \label{app:autoencoder_design}
This section details supplementary analysis of architectural configurations for the AE. The experimental protocol evaluates each AE architecture by first executing a GA phase (Sec.~\ref{sec:GA}) to measure the GA reconstruction loss (MSE) for a given communication reduction level $R$ (which determines $d_z$). Following this alignment, each AE is integrated into DFT with $N=1$ client to assess its impact on downstream top-1 accuracy.

\subsubsection{Activation Functions in AE architectures} \label{app:activations}
\begin{figure}[!h]
    \centering \small
    \includegraphics[width=1\linewidth]{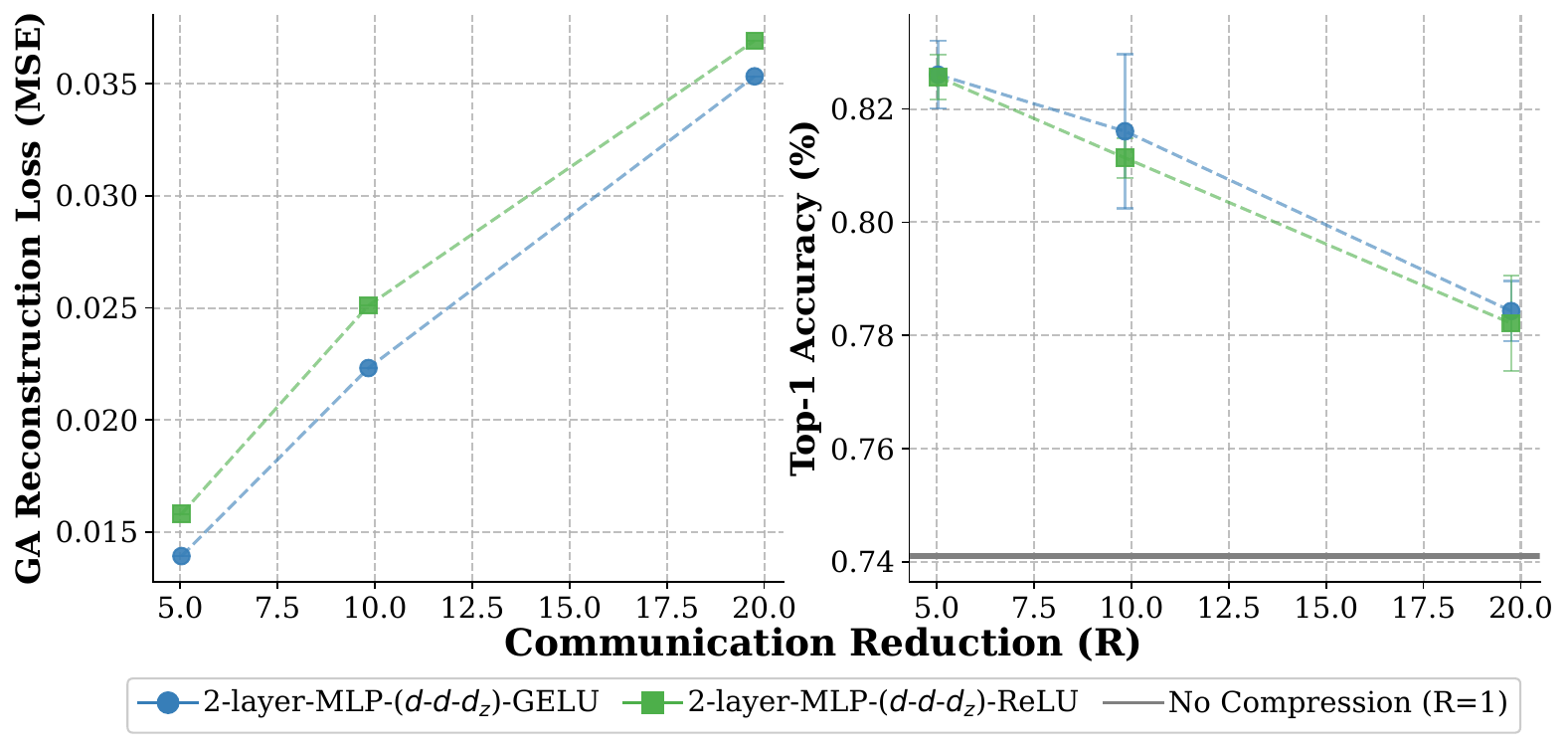}
    \caption{Evaluation of AE architectures varying communication reduction levels ($R$) on Food101 ($N=1$). Left: GA reconstruction loss (MSE) Right: Corresponding DFT top-1 accuracy.}
    \label{fig:activation_functions}
\end{figure}

To justify the selection of the activation function, we compare two architecturally identical models differing only in their non-linearity: ReLU and GELU. As illustrated in Fig.~\ref{fig:activation_functions}, the GELU-based architecture consistently yields a lower GA reconstruction loss. Regarding downstream top-1 accuracy, GELU maintains a marginal advantage, albeit with slightly higher variance. Given its superior reconstruction fidelity and minor improvement in task performance, GELU is adopted across all AE architectures in our primary experiments.

\subsubsection{AE Parameter vs. Accuracy Trade-off.} \label{sec:app_ae_param}
For the majority of evaluated AE architectures, the additional  computational overhead remains marginal relative to the client-side model ($<2\%$ parameter increase), assuming a split layer of $s=5$ for ViT-B/32. Nevertheless, to accommodate edge scenarios with stricter computational constraints, we analyze the compression performance of various AE architectures as a function of their parameter count. The parameter count for a given architecture scales dynamically, as the latent dimension $d_z$ dictates the size of the final hidden layer.

\begin{figure}[!t]
    \centering \small
    \includegraphics[width=\linewidth]{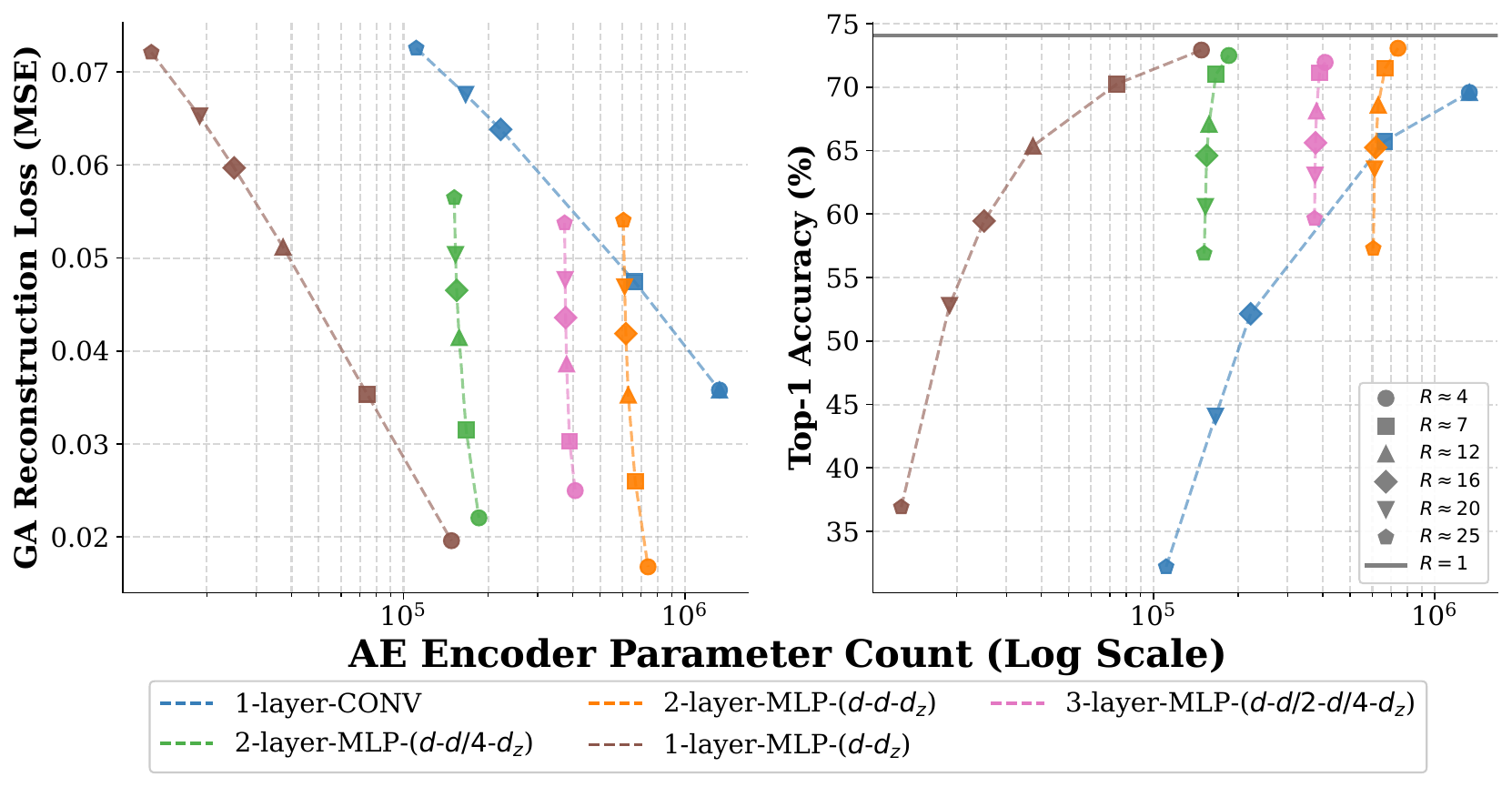}
    \caption{Evaluation of encoder parameter counts for AE architectures across varying communication reduction levels ($R$) on the Food101 dataset ($N=1$). Only the encoder parameter count is denoted, as client-side computation is our primary focus; the decoder is symmetric and of equal size. Left: GA reconstruction loss (MSE). Right: Corresponding DFT top-1 accuracy.}
    \label{fig:ae_comp_eff}
\end{figure}

As depicted in Fig.~\ref{fig:ae_comp_eff}, the $2\text{-layer-MLP-}(d\text{-}d\text{-}d_z)$ configuration sustains high accuracy across most compression ratios. The deeper $3\text{-layer-MLP-}(d\text{-}d/2\text{-}d/4\text{-}d_z)$ variant demonstrates improved performance specifically under extreme compression ($R\approx25$). Notably, increasing the parameter count does not strictly correlate with improved downstream task accuracy; however, the expanded model capacity does reduce reconstruction MSE at higher compression levels. Conversely, highly parameterized architectures yield diminishing returns at lower compression ratios. As such, the $2\text{-layer-MLP-}(d\text{-}d\text{-}d_z)$ structure offers an optimal trade-off for maximizing fine-tuning accuracy, introducing only a $1.6\%$ increase in client-side parameters. For environments governed by strict computational limitations, the $2\text{-layer-MLP-}(d\text{-}d/4\text{-}d_z)$ architecture provides viable performance while reducing the parameter count by an order of magnitude.


\newpage
\subsection{Impact of \cls\ Token Compression.} \label{app:cls}
The \cls\ token aggregates global contextual information that is critical for downstream classification at the server-side layers. Consequently, maintaining its representation fidelity is vital for model convergence. To quantify the impact of compressing this feature, we evaluate the top-1 accuracy under fixed communication reduction levels ($R$). To ensure a rigorous comparison, when the \cls\ token is compressed, the latent dimension $d_z$ of the bottleneck is increased proportionally to match the exact communication reduction level of the uncompressed-\cls\ baseline.

As demonstrated in Table \ref{tab:cls}, subjecting the \cls\ token to compression induces a severe drop in downstream task performance. This degradation is highly pronounced across all communication reduction levels and further degrades as the reduction ratio increases. These findings validate our framework design choice to transmit the \cls\ token uncompressed, prioritizing its semantic integrity to stabilize fine-tuning over edge networks. While preserving the \cls\ token intact guarantees training stability, it leaves the framework reliant on this specific architectural design choice.

\begin{table}[!h]
\centering \small
\caption{Impact of \cls\ token compression on DFT top-1 accuracy across degrees of communication reduction ($R$).}
\label{tab:cls}
\begin{tabular}{lccc}
\toprule
\textbf{Compress \cls} & \textbf{Low ($R\approx5$)} & \textbf{Mid ($R\approx10$)} & \textbf{High ($R\approx20$)} \\
\midrule
False (\method) & \textbf{82.4 $\pm$ 0.5} & \textbf{82.0 $\pm$ 0.5} & \textbf{79.0 $\pm$ 0.8} \\
True            & 73.9 $\pm$ 4.5          & 62.6 $\pm$ 2.5          & 46.8 $\pm$ 31.1 \\
\bottomrule
\end{tabular}
\end{table}

\end{document}